\documentclass[aps,prl,twocolumn,groupedaddress,preprintnumbers,]{revtex4-2}

\usepackage{amsmath}
\usepackage{amssymb}
\usepackage{amsfonts}
\usepackage[dvipsnames]{xcolor}
\usepackage{hyperref}
\hypersetup{
	pdftex,
	pdfstartview=FitH,
	bookmarksnumbered=true,
	pagebackref,
	colorlinks,
	breaklinks,
	urlcolor=Maroon,
	linkcolor=Maroon,
	citecolor=Bittersweet
}

\usepackage{graphicx}
\DeclareGraphicsExtensions{.pdf,.png,.jpg,.mps}
\usepackage{tikz}
\usetikzlibrary{matrix,decorations.pathmorphing,decorations.markings,arrows,positioning,intersections,mindmap,backgrounds}
\usetikzlibrary{arrows.meta,fit,shadows}
\usetikzlibrary{positioning,calc} 

\usepackage{booktabs}
\usepackage{physics}

\usepackage{array}
\usepackage{diagbox}

\usepackage{lmodern} 
\usepackage{scalerel,accents,trimclip}
\usepackage{CJK}

\usepackage{shuffle}
\usepackage{cleveref}
\usepackage{subfig}
\usepackage{float}  
\usepackage{ragged2e}

\newcommand{\zb}{\bar{z}}

\newcommand{\stageCorrect}{\textcolor{green!50!black}{$\checkmark$}}
\newcommand{\stageWrong}{\textcolor{red!80!black}{$\times$}}

\usepackage{graphicx}

\begin{document}

\preprint{MIT-CTP 6124}
	
\begin{CJK*}{UTF8}{gbsn}

\title{The Back-to-Back Plateau of the Energy-Energy Correlator\\ from Weak to Strong Coupling}

    \author{Hao Chen (陈豪)$^{1}$}
    \email{hao\_chen@mit.edu}
	\author{Ian Moult$^{2}$}
    \email{ian.moult@yale.edu}
    \author{Bo Wang (王波)$^{2}$}
	\email{bo.x.wang@yale.edu}
    \author{Hua Xing Zhu (朱华星)$^{3}$}
	\email{zhuhx@pku.edu.cn}
    \affiliation{{$^1$}Center for Theoretical Physics - a Leinweber Institute,\\
    Massachusetts Institute of Technology, Cambridge, MA 02139, USA}
	\affiliation{{$^2$}Department of Physics, Yale University, New Haven, CT 06511}
    \affiliation{{$^3$}School of Physics, Peking University, Beijing 100871, China}

	\date{\today}

	\begin{abstract} The back-to-back limit of the energy-energy correlator (EEC) exhibits a finite plateau, capturing the evolution from perturbative soft-collinear radiation at weak coupling to string dynamics at strong coupling. At weak coupling, we resum its leading and next-to-leading-power logarithms through N$^3$LL in $\mathcal N=4$ SYM, including non-planar contributions. Combining this weak-coupling endpoint with bootstrap and strong-coupling data, we construct an interpolation for the plateau height from weak to strong coupling. The calculation uses a human-in-the-loop agentic workflow.
	\end{abstract}

	\maketitle
	
	\end{CJK*}

\noindent{\bf Introduction.} Understanding how an observable evolves as the coupling is varied is a central problem in quantum field theory. Energy correlators make this question concrete by measuring angular correlations of energy flow~\cite{Sterman:1975xv,Basham:1978bw,Basham:1977iq}. In Quantum Chromodynamics (QCD), recent measurements inside jets resolve the transition from perturbative parton showering to hadronization as the angular scale is varied~\cite{Komiske:2022enw,ALICE:2024dfl,Electron-PositronAlliance:2025fhk}, motivating new descriptions of energy correlations in the hadronic regime~\cite{Lee:2025okn,Guo:2025zwb,Chang:2025kgq,Kang:2025zto}. A related one-point correlator follows the change from quark to pion degrees of freedom in an experimentally accessible angular distribution, as the collision energy flows from a fermionic ultraviolet regime to a bosonic infrared regime~\cite{Riembau:2025wjc}. However, a first-principles description through this crossover is difficult due to the running coupling and confinement. 

In $\mathcal N=4$ supersymmetric Yang--Mills (SYM) theory, the coupling can be varied without introducing a mass scale. The theory can therefore be studied within the conformal-collider setup~\cite{Hofman:2008ar}.
Integrability has determined the cusp anomalous dimension across
coupling~\cite{Beisert:2006ez,Basso:2007wd} and followed the Konishi dimension from perturbation theory to its short-string regime in AdS$_5\times$S$^5$~\cite{Gromov:2009zb}. More generally, the quantum spectral curve formulates the planar spectral problem at finite coupling~\cite{Gromov:2013pga}. The next question is how this spectral information enters dynamical observables. Finite-coupling studies include amplitudes in special kinematic limits~\cite{Basso:2020xts} and scattering on the Coulomb branch~\cite{Alday:2025pmg}. For energy flow, the light-transform relation to local four-point functions~\cite{Belitsky:2013xxa,Belitsky:2013bja,Belitsky:2013ofa} has led to conformal collider bootstrap bounds on the energy-energy correlator (EEC)~\cite{Dempsey:2025yiv}.

We focus on the back-to-back regime of the EEC. At weak coupling, nearly antipodal energetic particles act as lightlike color sources, and Wilson lines describe their leading-power soft radiation. The associated Sudakov logarithms admit a flux-tube interpretation, with the cusp anomalous dimension giving the flux energy density~\cite{Alday:2007mf}. The large-spin trajectory has a long-string dual at strong coupling~\cite{Gubser:2002tv}; a related emergence of string-like flux-tube excitations was found for a static quark--antiquark pair~\cite{Klebanov:2006jj}. We ask how these gauge theories and string descriptions are reflected in inclusive energy flow. Sudakov resummation turns the fixed-order singularity of the EEC into a finite plateau at fixed positive coupling~\cite{Belitsky:2013ofa,Korchemsky:2019nzm,Dokshitzer:1999sh}. Its height has an essential singularity at zero coupling that does not admit weak-coupling power expansion, while at strong coupling it approaches the supergravity prediction. At weak coupling, the plateau height is dominated by the leading-power (LP) resummation whose contribution is controlled by the twist-two operators. The resummed logarithms at next-to-leading-power (NLP) corrects the plateau height and supply weak-coupling information beyond the leading-power contribution. These logarithms can be obtained using large-spin methods and crossing symmetry~\cite{Alday:2015eya,Alday:2016njk,Henriksson:2017eej,Chen:2023wah,Chen:2023amz}, but extending the calculation to high logarithmic accuracy requires extensive symbolic computation. In the plateau limit, the NLP contribution also produces non-analytic behavior in coupling, which is numerically consistent with the EEC bootstrap prediction~\cite{Dempsey:2025yiv}.

In this Letter, we resum the leading- and next-to-leading-power logarithms of the back-to-back EEC through N$^3$LL accuracy in $\mathcal N=4$ SYM, including nonplanar contributions. The NLP result contains both the usual LP Sudakov structure and an error function generated by the all-order logarithmic sum. We then combine the resulting weak-coupling endpoint with finite-coupling bootstrap information and the strong-coupling result to construct an asymptotically constrained interpolation for the plateau height. The calculation was implemented using a human-in-the-loop agentic workflow. An audit of the research trajectory and an independent test show that agents perform many local calculations reliably but still require substantial intervention from human experts in long-horizon calculations.

\vspace{0.8em}

\noindent{\bf Energy-energy correlator.} We focus on a process generated by local twist-2 half-BPS operators $\mathcal{O}(x_i)$ injecting with a total momentum $q$. EEC encodes the correlation between energy fluxes deposited by a pair of detectors and can be written as
\begin{equation*}
    \text{EEC}(z) =  \frac{8 \pi^2}{q^2 \sigma_0} \int \mathrm{d}^4 x_{13} \, e^{iq\cdot  x_{13}} \langle \mathcal{O} (x_1)\mathcal{E}(n_2)\mathcal{E}(n_4)\mathcal{O}(x_3)\rangle \, ,
\end{equation*}
where $\sigma_0$ is the total cross-section, $x_{13}=x_1-x_3$. The energy flow operator $\mathcal{E}(n_i)$ is constructed by integrating the stress-tensor $T_{\mu \nu}$ along the unit direction $n=(1,\hat{n})$~\cite{Korchemsky:1997sy,Sveshnikov:1995vi},
\begin{equation} 
\mathcal{E}(\vec{n}) = \lim_{r \to \infty} r^2 \int_{0}^{\infty} dt \, {n}_i\, T_{0i}(t, r\hat{n}) \,. 
\end{equation}
In the rest frame of the scalar source, rotational invariance reduces the function dependence of EEC to the relative detector angle $\theta$, or $z=\sin^2(\theta/2)$.

In the back-to-back limit, $y=1-z\to0$, soft and collinear radiation
generates logarithmically enhanced terms at leading and next-to-leading
power of EEC
$$
\text{EEC}(y) \sim \sum_{n=1}^{\infty} \sum_{m=0}^{2n-1} a^n \left( c_{n,m}\frac{\log^m y}{y} + d_{n,m} \log^my\right) \, ,
$$
where $a$ is the coupling defined as $a=\lambda/(4\pi^2)$. The $y^{-1}$ and $y^0$ terms define the LP and NLP contributions, respectively. Fixed-order perturbation theory breaks down when $aL_y^2$ becomes of order one, where $L_y=\log y$. At LP, these logarithms resum into the impact-parameter representation~\cite{Belitsky:2013ofa,Korchemsky:2019nzm}
\begin{align*}
    &{\rm EEC}_{\rm LP}(y;a) = \frac{H(a)}{4y} \int _0^{
    \infty}db \, b \,J_0(b)  \\
    &\quad \times \exp{ - \frac{1}{2} \Gamma_{\bf cusp} (a) \log^2\left(\frac{b^2}{yb^2_0}\right) -\Gamma_{\bf coll}(a) \log\left(\frac{b^2}{yb^2_0}\right) } .
\end{align*}
where $b_0 = 2e^{-\gamma_E}$ and $\gamma_E$ are the Euler-Mascheroni constant. The physical data $\Gamma_{\bf cusp}$, $\Gamma_{\bf coll}$ and hard function $H(a)$ admit the weak-coupling expansion $\Gamma_{\bf cusp} = a +\mathcal{O}(a^2)$~\cite{Kotikov:2004er,Beisert:2006ez,Freyhult:2007pz,Grozin:2015kna,Henn:2019swt}, $\Gamma_{\bf coll} = -\frac{3}{2}  a^2 \zeta_3 +\mathcal{O}(a^3)$~\cite{Kotikov:2004er,Freyhult:2009my} and $H(a) = 1 -\frac{\pi^2}{6}a+\mathcal O(a^2)$~\cite{Eden:2012rr,Alday:2013cwa}. Setting $x=b^2/(yb_0^2)$ gives
$b\,db=(yb_0^2/2)\,dx$, canceling the explicit $1/y$ factor.
As $y\to0$, the Bessel function tends to one, and the remaining
integral is Gaussian in $\log x$ we have~\cite{Dempsey:2025yiv}
\begin{align*}
P_{\rm LP}(a)
&\equiv\lim_{y\to0}{\rm EEC}_{\rm LP}(y;a)
\nonumber\\
&=
\sqrt{\frac{\pi}{2}}\,e^{-2\gamma_E}
\frac{H(a)}{\sqrt{\Gamma_{\rm cusp}(a)}}
\exp\left[
\frac{\bigl(1-\Gamma_{\rm coll}(a)\bigr)^2}
{2\Gamma_{\rm cusp}(a)}
\right].
\end{align*}
The LP result is therefore finite at fixed small coupling, but its weak-coupling behavior, $P_{\rm LP}(a)\sim a^{-1/2}e^{1/(2a)}$, is non-analytic at $a=0$ cannot be obtained at any finite perturbative order.

The leading-power result $P_{\rm LP}(a)$ does not, by itself, determine the full endpoint. We therefore define the height of the endpoint plateau as a function of coupling $a$,
\begin{equation}
    P(a)\equiv\lim_{y\to0}{\rm EEC}(y;a)=P_{\rm LP}(a)+H_0(a)
\end{equation}
where $H_0(a)$ denotes the additional endpoint information not captured by the twist-two LP contribution. Its determination is needed to obtain the full plateau from $P_{\rm LP}(a)$ Ref.~\cite{Dempsey:2025yiv} numerically estimates the value of $H_0(a)$ at finite coupling from EEC boostrap data. The NLP resummation computed below provides weak-coupling approximation for this residual contribution.

\begin{figure}[t]
    \centering
    \includegraphics[width=\linewidth]{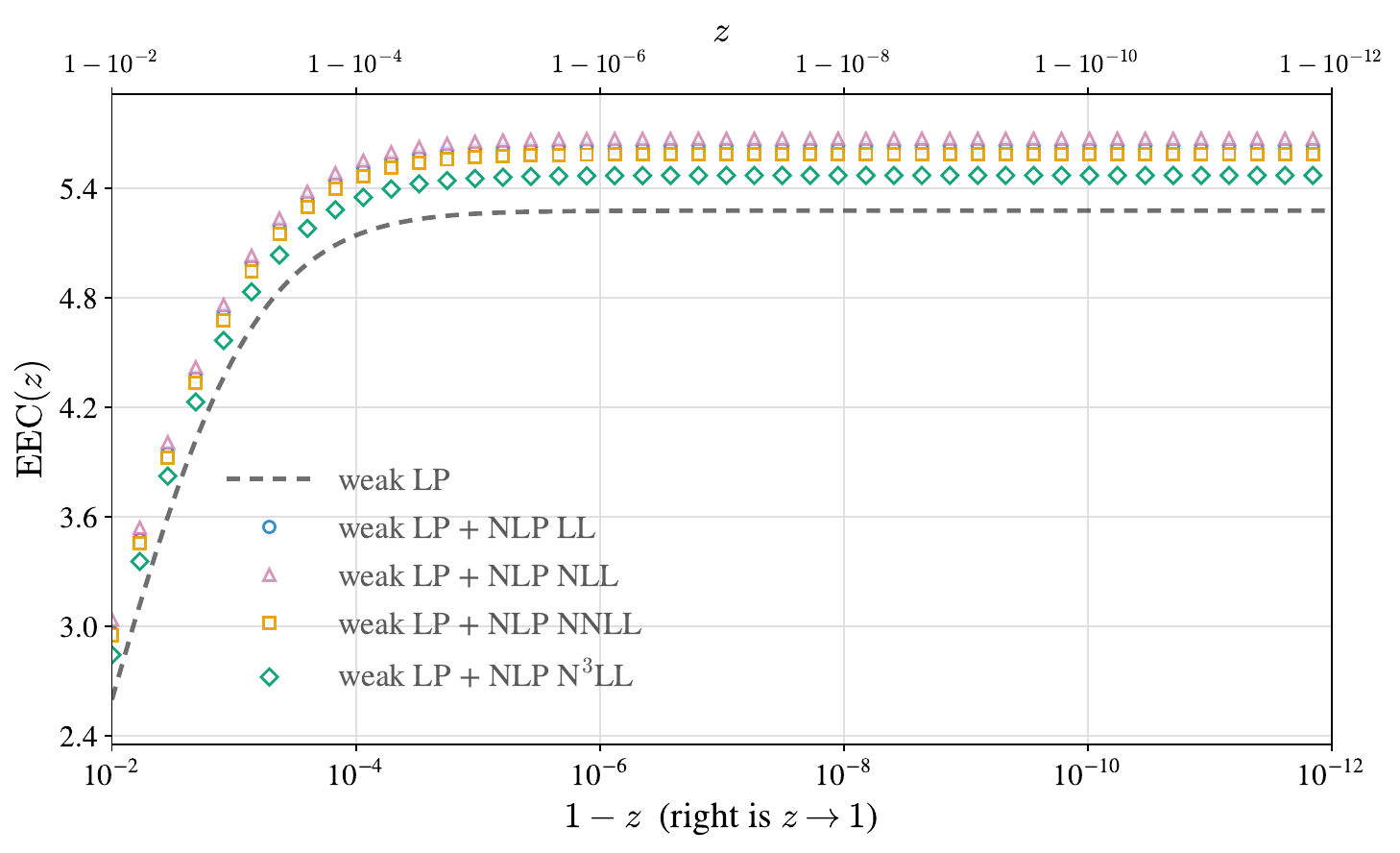}
    \caption{\justifying 
    Back-to-back EEC at $a=0.3$. The LP curve is obtained from the Bessel representation using the weak-coupling data of \cite[eq 5.14]{Dempsey:2025yiv}. The other curves add the NLP logarithmic towers through LL, NLL, NNLL, and N$^3$LL accuracy in the planar limit. The flattening as $z\to1$ indicate the finite Sudakov plateau obtained after resumming the endpoint logarithms.}
    \label{fig:eec-plateau-window}
\end{figure}

\begin{figure*}[t!]
    \centering
    \includegraphics[width=1\linewidth]{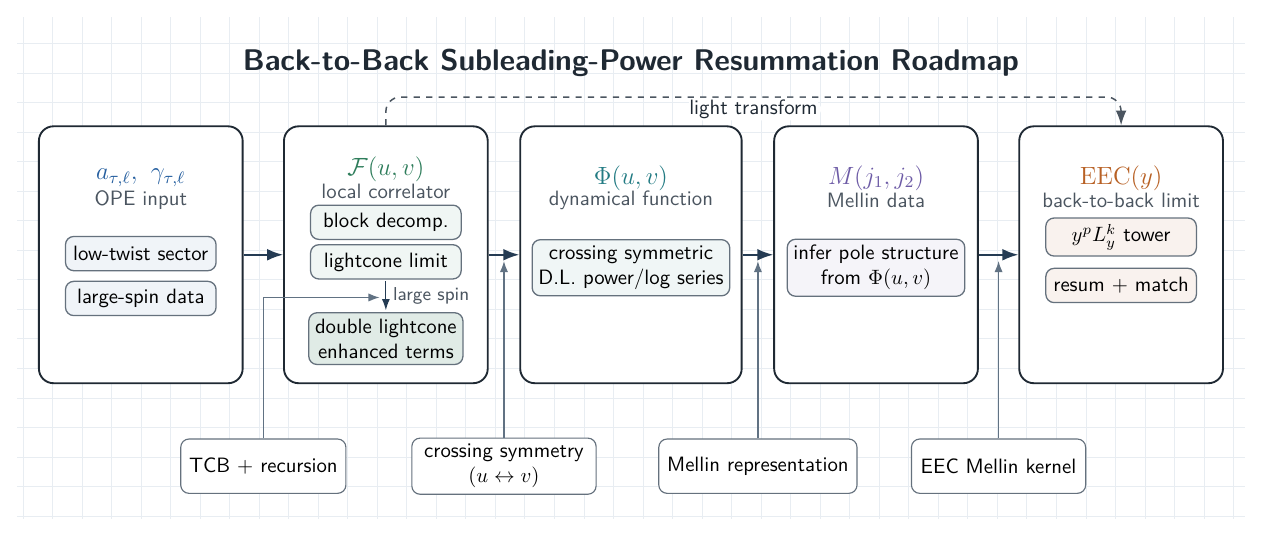}
    \caption{\justifying
    Subleading power back-to-back resummation chain. The calculation starts from twist-2 block data, passes through large-spin and TCB resummation, and ends with the light transform to the EEC LP and NLP logarithmic towers via Mellin representation. Fixed-order EEC determine the boundary coefficients.}
    \label{fig:b2b-flowchart}
\end{figure*}

At weak coupling, the essential singularity of $P_{\rm LP}(a)$ makes the twist-two Sudakov contribution dominant. At strong coupling, the full endpoint approaches the supergravity prediction. We will show $P(a)$ to compare these two limits, and one can see the evolution explicitly.

\vspace{0.8em}

\noindent{\bf Algorithm and agentic workflow.} To determine the endpoint beyond the leading-power contribution, we follow the large-spin method developed in~\cite{Chen:2023wah,Chen:2023amz}. The required inputs are the twist-two anomalous dimension~\cite{Kotikov:2004er,Beisert:2006ez,Kotikov:2007cy,Bajnok:2008qj,Beccaria:2009vt,Kniehl:2021ysp,Kniehl:2024tvd}, the corresponding OPE coefficients~\cite{Eden:2012rr}, and fixed-order result of EEC~\cite{Belitsky:2013ofa, Henn:2019gkr}. Starting from the superconformal-block decomposition of the local four-point function, we expand the correlator in the double-lightcone limit and reorganize the direct spin sums in a finite-difference large-spin basis. The resulting asymptotic data are resummed in terms of twist conformal blocks (TCB), after which crossing symmetry reconstructs the singular part of the correlator $\Phi(u,v)$ that contributes to singular behaviors in EEC. Based on the Mellin-space approach, light transform then maps these singularities to the LP and NLP logarithmic towers of the back-to-back EEC. The generic all-order coefficients are determined by the large-spin data, while the remaining exceptional boundary coefficients are fixed by matching to the known fixed-order endpoint expansion. The underlying algorithm is shown in~\Cref{fig:b2b-flowchart}.

Although the method is established, its implementation at N$^3$LL requires extensive symbolic manipulation and careful truncation at each step~\Cref{fig:b2b-flowchart}. We used agents to generate and test these implementations with human supervision. The Planner supplied each task with its input, target output, and prescribed checks. Two executor routes produced separate implementations, tested against algebraic identities, large-spin expansion, crossing symmetry, Mellin residues and fixed-order coefficients. The human experts fixed the required expansion depth, handled exceptional terms, and reviewed the checks. Further workflow details are given in the Supplemental Material.

Our result for the back-to-back EEC, through next-to-leading power in $y$ and N$^3$LL logarithmic accuracy, can be organized as
\begin{align}
    {\rm EEC} (y) =  \frac{1}{y} \mathcal G_{y} F(L_y,a) +G(L_y,a) + \mathcal{O}(y)\, ,
\end{align}
where $\mathcal G_y=e^{-aL^2_y/2}$ is the Gaussian Sudakov factor. 
Expanding the LP term $\mathcal G_y F(L_y,a)/y$ reproduces the known LP logarithmic coefficients through N$^3$LL. Since independent finite-coupling information from integrability~\cite{Beisert:2006ez,Freyhult:2009my} is already available for the twist-two LP sector, the central new perturbative result of this Letter is the NLP function $G(L_y,a)$. We give $F$ in the Supplemental Material and focus here on the NLP function $G$, shown in~\Cref{eq:NLPfunG}. We define $\mathcal R_y=\mathrm{erf}(\sqrt a L_y/\sqrt2)$ for convenience. For the endpoint plots, we retain the Bessel representation for the LP term and add the NLP expression in~\Cref{eq:NLPfunG}. 

\begin{figure*}
\begin{align}\label{eq:NLPfunG}
G(L_y,a)=&
\sqrt{\frac{\pi}{8}}\sqrt a
\left[-1+\left(1+\frac{\pi^2}{3}\right)a\right]\mathcal R_y
+\mathcal G_y\Bigg\{
-\frac{a}{6}-\frac{a}{2}L_y
+a^2\left[
-\frac14+\frac{7\pi^2}{48}+\frac{\zeta_3}{12}
+\frac{\pi^2}{6}L_y+\frac{7}{24}L_y^2
\right]
\nonumber\\
&\hspace{0.3cm}
+a^3\left[
-\frac{\pi^2}{48}L_y^3
-\left(\frac{11\pi^2}{96}+\frac{41\zeta_3}{24}\right)L_y^2
\right]
+a^4\left(\frac{7\pi^2}{576}+\frac{\zeta_3}{3}\right)L_y^4
-\frac{a^2}{N_c^2}\left(\frac{\pi^2}{4}+3\zeta_3\right)
-\frac{a^3}{N_c^2}\left(\frac{\pi^2}{8}+\frac{3\zeta_3}{2}\right)L_y^2
\Bigg\}
\nonumber\\
&+\frac{a}{6}
+a^2\left[
-\frac54-\frac{3\pi^2}{16}
+\frac{3\pi^2}{4}\log2
+\frac{13\zeta_3}{24}
+\frac{1}{N_c^2}\left(\frac{\pi^2}{4}+3\zeta_3\right)
\right] + \cdots.
\end{align}
\end{figure*}

At LP, all logarithmic towers multiply the Gaussian Sudakov factor $\mathcal G_y$. The NLP result contains terms proportional to $\mathcal E_y$ and $\mathcal R_y$, together with a term independent of $L_y$. The error function sums an infinite tower of perturbative logarithms. As $y\to0$, the terms multiplying $\mathcal G_y$ vanish and $\mathcal R_y\to-1$, leaving a finite NLP correction to the plateau height. \Cref{fig:eec-plateau-window} shows the result at $a=0.3$ with the LL through N$^3$LL towers. The plateau captures the contributions from different operators, thereby providing a direct observable for non-perturbative dynamics.

\vspace{0.8em}

\begin{figure*}[t!]
    \centering
    \includegraphics[width=0.95\textwidth]{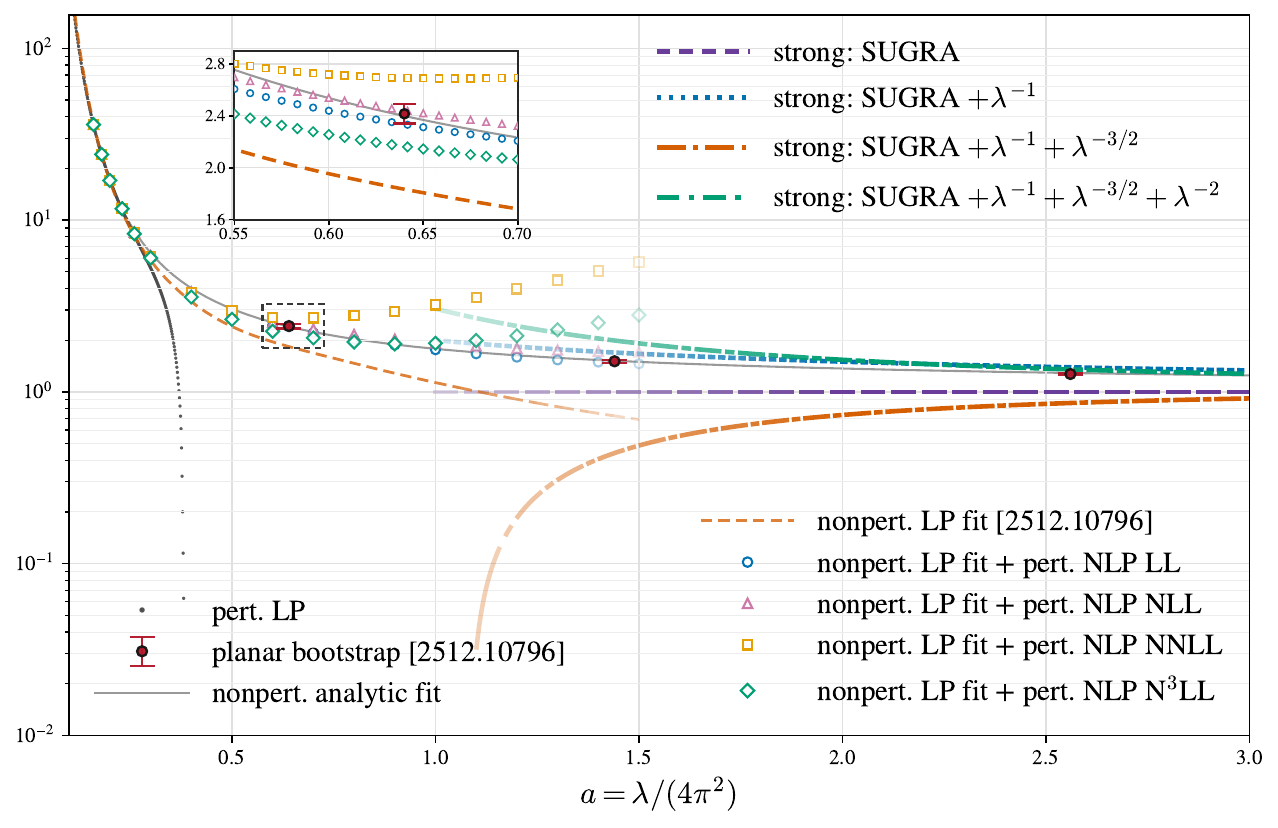}
    \caption{\justifying
Endpoint plateau height $P(a)=\lim_{y\to0}{\rm EEC}(y;a)$ as a function of $a=\lambda/(4\pi^2)$in planar limit. The weak-coupling LP dotted curve is obtained from the Bessel representation. The orange dashed curve is the finite-coupling twist-two LP reconstruction of Ref.~\cite{Dempsey:2025yiv}. The additive endpoint term $H_0(a)$, which contains contributions not fixed by that reconstruction, is not included. The markers add the perturbative NLP towers through LL, NLL, NNLL, and N$^3$LL to the finite-coupling LP reconstruction. The planar-bootstrap points~\cite{Dempsey:2025yiv} shown as points with uncertainties constrain the full plateau. The strong-coupling curves show the supergravity result and corrections from stringy effects. The solid gray curve is the non-perturbative fit in~\Cref{eq:finite_plateau}, constructed from the weak and strong coupling data.}
    \label{fig:eec-plateau-coupling}
\end{figure*}

\noindent{\bf Endpoint plateau in general coupling.} Now let us turn to discuss the endpoint plateau of EEC in general coupling. As we already mentioned, the NLP result supplements the weak-coupling LP plateau. On the contrary side, at strong coupling, the endpoint plateau admits the asymptotic expansion~\cite{Ren:2026zxs,Jackson:2026eml},
\begin{align*}
P_{\rm strong}&(a)=1+\frac{1}{a}+\frac{\pi ^2-120 \zeta (3)}{2 \pi ^3 a^{3/2}} \\
&\hspace{-2em}+\frac{\frac{30 \zeta (3)+680}{\pi ^2}-\frac{30 (239 \zeta (3)+10 \zeta (5))}{\pi ^4}+1321}{120 a^2} + \mathcal{O}(a^{-\frac{5}{2}}) \, ,
\end{align*}
where the leading constant $1$ is the supergravity contribution, and the inverse powers of $a$ encode stringy corrections. One would see that the two limits have quiet different analytic forms,
\begin{align*}
P(a)\sim  \begin{cases}
    a^{-1/2}e^{1/(2a)},\qquad & a\to0, \\
    1+\frac{1}{a}+\mathcal O(a^{-3/2}),\qquad & a\to\infty.
\end{cases}
\end{align*}
At weak coupling it is impossible to reproduce this non-analytic behavior in any perturbative expansion. To connect these limits, we construct an asymptotically constrained interpolation for the full plateau,
\begin{align}\label{eq:finite_plateau}
P(a)=&
1+
\sqrt{\frac{\pi}{2}}\,
e^{-2\gamma_E+\pi^2/24}
\frac{e^{1/(2a)}}{\sqrt a} \nonumber \\
& \quad \times
\frac{1+p\sqrt a}
{1+q\sqrt a+
\sqrt{\frac{\pi}{2}}e^{-2\gamma_E+\pi^2/24}p\,a},  
\end{align}
with $p=2.34 $ and $q=1.81$. By construction, this form reproduces the leading non-analytic Sudakov behavior at weak coupling and matches the supergravity result with the leading $1/a$ correction at strong coupling. The curve with
$p=2.34$ and $q=1.81$ is compared with the bootstrap data in~\Cref{fig:eec-plateau-coupling}. It should be regarded as a compact interpolation of the available all-coupling information, but it fits the finite-coupling bootstrap data well.

\Cref{fig:eec-plateau-coupling} summarizes these complementary descriptions. The twist-two LP curve dominates in the weak-coupling regime and the NLP logarithmic towers derived in this Letter provide new weak-coupling information beyond the LP reconstruction, modifying both the endpoint height and the manner in which the plateau is approached. As expected, the perturbative description becomes increasingly accurate as the coupling decreases, while deviations from the finite-coupling interpolation become visible when the coupling is increased beyond the regime where the weak-coupling expansion provides quantitative control. This expectation is consistent with non-perturbative localization results for several observables in $\mathcal N=4$ SYM, where singularities in the exact coupling dependence appear at $\lambda\sim\pi^2$, corresponding to $a\sim1/4$ in our convention. Including NLP corrections extends the perturbative description, but does not remove the eventual breakdown of the weak coupling expansion due to higher-order and nonperturbative effects.

On the other hand, the strong coupling description is controlled by the stringy resummation of the endpoint dynamics, which stabilizes the plateau and provides an accurate description from parametrically large coupling down to intermediate coupling ($a\sim 0.5$). Therefore, the combination of weak-coupling resummation and strong-coupling string theory leaves a finite intermediate-coupling window where the nonperturbative structure of the endpoint plateau remains to be understood.

The analytic interpolation shown in~\Cref{fig:eec-plateau-coupling} provides a prediction for this crossover region, constrained by the weak- and strong-coupling asymptotics as well as available finite-coupling data. Further constraints from the planar bootstrap on the endpoint plateau in this intermediate regime would provide a direct test of this interpolation and shed light on the universal structure of the back-to-back EEC across coupling. In conclusion, the endpoint plateau provides a unified quantity on which perturbative resummation, conformal bootstrap data, and stringy corrections can be directly compared.

\begin{table*}[t]
\centering
\scriptsize
\caption{\justifying
Step-by-step tests for the seven models. A check mark indicates that the model obtained the correct result for that step, while a cross indicates an incorrect or incomplete result. The last row gives the total number of correct steps.}
\setlength{\tabcolsep}{3.0pt}
\renewcommand{\arraystretch}{1.10}
\begin{tabular}{@{}clcccccccc@{}}
\toprule
Step & Operation & GPT-5.5 &  Gemini 3.5 Flash & Qwen3.7-Max & DeepSeek V4-Pro & Kimi K2.6 & GLM-5.1 & Seed2.1 Turbo \\
\midrule
1  & Block expansion            & \stageCorrect  & \stageCorrect & \stageCorrect & \stageCorrect & \stageCorrect & \stageCorrect & \stageCorrect \\
2  & OPE convolution            & \stageCorrect  & \stageCorrect & \stageCorrect & \stageCorrect & \stageWrong   & \stageCorrect & \stageCorrect \\
3  & Integration by parts       & \stageCorrect  & \stageCorrect & \stageCorrect & \stageCorrect & \stageCorrect & \stageCorrect & \stageCorrect \\
4  & Finite differences         & \stageCorrect  & \stageCorrect & \stageCorrect & \stageCorrect & \stageCorrect & \stageCorrect & \stageCorrect \\
5  & Data extraction            & \stageWrong    & \stageCorrect & \stageWrong   & \stageWrong   & \stageWrong   & \stageWrong   & \stageWrong   \\
6  & Large-spin expansion       & \stageWrong    & \stageWrong   & \stageWrong   & \stageWrong   & \stageWrong   & \stageWrong   & \stageWrong   \\
7  & Large-spin substitution    & \stageCorrect  & \stageCorrect & \stageCorrect & \stageWrong   & \stageCorrect & \stageWrong   & \stageWrong   \\
8  & Twist-block basis          & \stageCorrect  & \stageCorrect & \stageWrong   & \stageWrong   & \stageWrong   & \stageWrong   & \stageWrong   \\
9  & Shifted twist blocks       & \stageCorrect  & \stageWrong   & \stageCorrect & \stageWrong   & \stageWrong   & \stageWrong   & \stageWrong   \\
10 & Explicit twist-block sum   & \stageWrong    & \stageWrong   & \stageWrong   & \stageWrong   & \stageWrong   & \stageWrong   & \stageWrong   \\
11 & Correlator reconstruction  & \stageWrong    & \stageCorrect & \stageWrong   & \stageWrong   & \stageWrong   & \stageWrong   & \stageWrong   \\
12 & Mellin construction        & \stageCorrect  & \stageWrong   & \stageWrong   & \stageWrong   & \stageWrong   & \stageWrong   & \stageWrong   \\
13 & EEC projection             & \stageCorrect  & \stageWrong   & \stageWrong   & \stageWrong   & \stageWrong   & \stageWrong   & \stageWrong   \\
14 & Two-loop expansion         & \stageCorrect  & \stageCorrect & \stageCorrect & \stageCorrect & \stageWrong   & \stageWrong   & \stageWrong   \\
15 & Three-loop expansion       & \stageWrong    & \stageWrong   & \stageWrong   & \stageWrong   & \stageWrong   & \stageWrong   & \stageWrong   \\
\midrule
& Correct steps & $10/15$  & $9/15$ & $7/15$ & $5/15$ & $4/15$ & $4/15$ & $4/15$ \\
\bottomrule
\end{tabular}
\label{tab:step-by-step-results}
\end{table*}

\vspace{0.8em}

\noindent{\bf Human supervision and autonomous evaluation.} The calculation above was carried out with GPT-5.5 in the HITL workflow. We examine the expert interventions in this trajectory and test seven models on fifteen individual steps without human feedback. Each benchmark task starts independently from validated upstream input from the reference calculation, so errors do not propagate between tasks.

In Step~5, GPT-5.5 chose an expansion depth that is not sufficient for the next transformations. It omitted terms needed after crossing and Mellin projection. These omissions propagated through the initial reconstruction and produced incorrect endpoint coefficients. We corrected the truncation by working backward from the final N$^3$LL LP and NLP coefficients. The required expansion depth was thus set by the final observable, several transformations beyond the immediate task.

\Cref{tab:step-by-step-results} reports the independently initialized tests. The first four steps yielded $x$ correct outputs out of 28, compared with $x$ out of 77 for Steps~5--15. None of the seven models completed the large-spin expansion in Step~6 or the final fixed-order EEC comparison in Step~15, and no model passed all fifteen tasks. The later steps require more symbolic coding and involve changes between large-spin, twist-block, and Mellin representations. Their lower pass rates persist even with correct upstream input.

Token and cost totals include the whole computation, retries, and validation across the fifteen tasks. Several models consumed large token budgets while repeatedly pursuing the same incorrect construction. Several models spent large token budgets repeatedly pursuing the same incorrect construction. Models that passed more steps generally cost more, although their cost per correct stage varied.

\vspace{0.8em}

\noindent{\bf Discussion.} We have resummed the leading- and next-to-leading-power back-to-back logarithms through N$^3$LL, including nonplanar contributions. The NLP terms change both the plateau height and the approach to the endpoint. We also constructed an interpolation between the leading weak- and strong-coupling limits and compared it with finite-coupling bootstrap
data. Further bootstrap bounds at intermediate coupling would test this interpolation and help constrain the residual endpoint contribution
$H_0(a)$, whose full coupling dependence is not fixed by the logarithmic towers computed here.

The coupling dependence of the full plateau remains to be determined from first principles. Sharper planar-bootstrap bounds at intermediate coupling would test our interpolation and constrain $H_0(a)$ beyond the NLP logarithms computed here. Understanding this contribution would help explain how the operator contributions to the endpoint change between the weak-coupling Sudakov regime and the strong-coupling string regime. A further question is which features of the NLP resummation persist in QCD, where the coupling runs and hadronization becomes important at low scales. Agent-assisted calculations would help obtain the higher-order data needed to address these questions and extend the study to more complicated energy-flow observables.

\vspace{0.8em}

	\begin{acknowledgments}
	\noindent{\bf Acknowledgments.}
	    The authors would like to thank Yichao Tang for helpful discussions. HC is supported by the U.S. Department of Energy, Office of Science, Office of Nuclear Physics under grant Contract Number DE-SC0011090. IM and BW are supported by the DOE Early Career Award DE-SC0025581, the Sloan Foundation, and the Simons Collaboration on Confinement and QCD Strings. HXZ is supported by the National Natural Science Foundation of China under grant No. 12425505.
	\end{acknowledgments}

	\bibliography{refs}

\begin{thebibliography}{49}%
\makeatletter
\providecommand \@ifxundefined [1]{%
 \@ifx{#1\undefined}
}%
\providecommand \@ifnum [1]{%
 \ifnum #1\expandafter \@firstoftwo
 \else \expandafter \@secondoftwo
 \fi
}%
\providecommand \@ifx [1]{%
 \ifx #1\expandafter \@firstoftwo
 \else \expandafter \@secondoftwo
 \fi
}%
\providecommand \natexlab [1]{#1}%
\providecommand \enquote  [1]{``#1''}%
\providecommand \bibnamefont  [1]{#1}%
\providecommand \bibfnamefont [1]{#1}%
\providecommand \citenamefont [1]{#1}%
\providecommand \href@noop [0]{\@secondoftwo}%
\providecommand \href [0]{\begingroup \@sanitize@url \@href}%
\providecommand \@href[1]{\@@startlink{#1}\@@href}%
\providecommand \@@href[1]{\endgroup#1\@@endlink}%
\providecommand \@sanitize@url [0]{\catcode `\\12\catcode `\$12\catcode
  `\&12\catcode `\#12\catcode `\^12\catcode `\_12\catcode `\%12\relax}%
\providecommand \@@startlink[1]{}%
\providecommand \@@endlink[0]{}%
\providecommand \url  [0]{\begingroup\@sanitize@url \@url }%
\providecommand \@url [1]{\endgroup\@href {#1}{\urlprefix }}%
\providecommand \urlprefix  [0]{URL }%
\providecommand \Eprint [0]{\href }%
\providecommand \doibase [0]{https://doi.org/}%
\providecommand \selectlanguage [0]{\@gobble}%
\providecommand \bibinfo  [0]{\@secondoftwo}%
\providecommand \bibfield  [0]{\@secondoftwo}%
\providecommand \translation [1]{[#1]}%
\providecommand \BibitemOpen [0]{}%
\providecommand \bibitemStop [0]{}%
\providecommand \bibitemNoStop [0]{.\EOS\space}%
\providecommand \EOS [0]{\spacefactor3000\relax}%
\providecommand \BibitemShut  [1]{\csname bibitem#1\endcsname}%
\let\auto@bib@innerbib\@empty
\bibitem [{\citenamefont {Sterman}(1975)}]{Sterman:1975xv}%
  \BibitemOpen
  \bibfield  {author} {\bibinfo {author} {\bibfnamefont {G.~F.}\ \bibnamefont
  {Sterman}},\ }\bibfield  {title} {\bibinfo {title} {{Jet Structure in e+ e-
  Annihilation with Massless Hadrons}},\ }\href@noop {} {\  (\bibinfo {year}
  {1975})}\BibitemShut {NoStop}%
\bibitem [{\citenamefont {Basham}\ \emph
  {et~al.}(1978{\natexlab{a}})\citenamefont {Basham}, \citenamefont {Brown},
  \citenamefont {Ellis},\ and\ \citenamefont {Love}}]{Basham:1978bw}%
  \BibitemOpen
  \bibfield  {author} {\bibinfo {author} {\bibfnamefont {C.~L.}\ \bibnamefont
  {Basham}}, \bibinfo {author} {\bibfnamefont {L.~S.}\ \bibnamefont {Brown}},
  \bibinfo {author} {\bibfnamefont {S.~D.}\ \bibnamefont {Ellis}},\ and\
  \bibinfo {author} {\bibfnamefont {S.~T.}\ \bibnamefont {Love}},\ }\bibfield
  {title} {\bibinfo {title} {{Energy Correlations in electron - Positron
  Annihilation: Testing QCD}},\ }\href
  {https://doi.org/10.1103/PhysRevLett.41.1585} {\bibfield  {journal} {\bibinfo
   {journal} {Phys. Rev. Lett.}\ }\textbf {\bibinfo {volume} {41}},\ \bibinfo
  {pages} {1585} (\bibinfo {year} {1978}{\natexlab{a}})}\BibitemShut {NoStop}%
\bibitem [{\citenamefont {Basham}\ \emph
  {et~al.}(1978{\natexlab{b}})\citenamefont {Basham}, \citenamefont {Brown},
  \citenamefont {Ellis},\ and\ \citenamefont {Love}}]{Basham:1977iq}%
  \BibitemOpen
  \bibfield  {author} {\bibinfo {author} {\bibfnamefont {C.~L.}\ \bibnamefont
  {Basham}}, \bibinfo {author} {\bibfnamefont {L.~S.}\ \bibnamefont {Brown}},
  \bibinfo {author} {\bibfnamefont {S.~D.}\ \bibnamefont {Ellis}},\ and\
  \bibinfo {author} {\bibfnamefont {S.~T.}\ \bibnamefont {Love}},\ }\bibfield
  {title} {\bibinfo {title} {{Electron - Positron Annihilation Energy Pattern
  in Quantum Chromodynamics: Asymptotically Free Perturbation Theory}},\ }\href
  {https://doi.org/10.1103/PhysRevD.17.2298} {\bibfield  {journal} {\bibinfo
  {journal} {Phys. Rev. D}\ }\textbf {\bibinfo {volume} {17}},\ \bibinfo
  {pages} {2298} (\bibinfo {year} {1978}{\natexlab{b}})}\BibitemShut {NoStop}%
\bibitem [{\citenamefont {Komiske}\ \emph {et~al.}(2023)\citenamefont
  {Komiske}, \citenamefont {Moult}, \citenamefont {Thaler},\ and\ \citenamefont
  {Zhu}}]{Komiske:2022enw}%
  \BibitemOpen
  \bibfield  {author} {\bibinfo {author} {\bibfnamefont {P.~T.}\ \bibnamefont
  {Komiske}}, \bibinfo {author} {\bibfnamefont {I.}~\bibnamefont {Moult}},
  \bibinfo {author} {\bibfnamefont {J.}~\bibnamefont {Thaler}},\ and\ \bibinfo
  {author} {\bibfnamefont {H.~X.}\ \bibnamefont {Zhu}},\ }\bibfield  {title}
  {\bibinfo {title} {{Analyzing N-Point Energy Correlators inside Jets with CMS
  Open Data}},\ }\href {https://doi.org/10.1103/PhysRevLett.130.051901}
  {\bibfield  {journal} {\bibinfo  {journal} {Phys. Rev. Lett.}\ }\textbf
  {\bibinfo {volume} {130}},\ \bibinfo {pages} {051901} (\bibinfo {year}
  {2023})},\ \Eprint {https://arxiv.org/abs/2201.07800} {arXiv:2201.07800
  [hep-ph]} \BibitemShut {NoStop}%
\bibitem [{\citenamefont {Acharya}\ \emph {et~al.}(2026)\citenamefont {Acharya}
  \emph {et~al.}}]{ALICE:2024dfl}%
  \BibitemOpen
  \bibfield  {author} {\bibinfo {author} {\bibfnamefont {S.}~\bibnamefont
  {Acharya}} \emph {et~al.} (\bibinfo {collaboration} {ALICE}),\ }\bibfield
  {title} {\bibinfo {title} {{Exposing the parton-hadron transition within jets
  with energy-energy correlators in pp collisions at s=5.02{\,}{\,}TeV}},\
  }\href {https://doi.org/10.1103/tgtl-7xh9} {\bibfield  {journal} {\bibinfo
  {journal} {Phys. Rev. D}\ }\textbf {\bibinfo {volume} {114}},\ \bibinfo
  {pages} {032002} (\bibinfo {year} {2026})},\ \Eprint
  {https://arxiv.org/abs/2409.12687} {arXiv:2409.12687 [hep-ex]} \BibitemShut
  {NoStop}%
\bibitem [{\citenamefont {Bossi}\ \emph {et~al.}(2025)\citenamefont {Bossi}
  \emph {et~al.}}]{Electron-PositronAlliance:2025fhk}%
  \BibitemOpen
  \bibfield  {author} {\bibinfo {author} {\bibfnamefont {H.}~\bibnamefont
  {Bossi}} \emph {et~al.} (\bibinfo {collaboration} {Electron-Positron
  Alliance}),\ }\bibfield  {title} {\bibinfo {title} {{Energy Correlators from
  Partons to Hadrons: Unveiling the Dynamics of the Strong Interactions with
  Archival ALEPH Data}},\ }\href@noop {} {\  (\bibinfo {year} {2025})},\
  \Eprint {https://arxiv.org/abs/2511.00149} {arXiv:2511.00149 [hep-ph]}
  \BibitemShut {NoStop}%
\bibitem [{\citenamefont {Lee}\ and\ \citenamefont
  {Stewart}(2026)}]{Lee:2025okn}%
  \BibitemOpen
  \bibfield  {author} {\bibinfo {author} {\bibfnamefont {K.}~\bibnamefont
  {Lee}}\ and\ \bibinfo {author} {\bibfnamefont {I.~W.}\ \bibnamefont
  {Stewart}},\ }\bibfield  {title} {\bibinfo {title} {{Dihadron Fragmentation
  and the Confinement Transition in Energy Correlators}},\ }\href
  {https://doi.org/10.1103/m18j-xypt} {\bibfield  {journal} {\bibinfo
  {journal} {Phys. Rev. Lett.}\ }\textbf {\bibinfo {volume} {136}},\ \bibinfo
  {pages} {081902} (\bibinfo {year} {2026})},\ \Eprint
  {https://arxiv.org/abs/2507.11495} {arXiv:2507.11495 [hep-ph]} \BibitemShut
  {NoStop}%
\bibitem [{\citenamefont {Guo}\ \emph {et~al.}(2026)\citenamefont {Guo},
  \citenamefont {Yuan},\ and\ \citenamefont {Zhao}}]{Guo:2025zwb}%
  \BibitemOpen
  \bibfield  {author} {\bibinfo {author} {\bibfnamefont {Y.}~\bibnamefont
  {Guo}}, \bibinfo {author} {\bibfnamefont {F.}~\bibnamefont {Yuan}},\ and\
  \bibinfo {author} {\bibfnamefont {W.}~\bibnamefont {Zhao}},\ }\bibfield
  {title} {\bibinfo {title} {{Factorization and Resummation for the Nearside
  Energy-Energy Correlators}},\ }\href {https://doi.org/10.1103/4qkq-x5st}
  {\bibfield  {journal} {\bibinfo  {journal} {Phys. Rev. Lett.}\ }\textbf
  {\bibinfo {volume} {136}},\ \bibinfo {pages} {081904} (\bibinfo {year}
  {2026})},\ \Eprint {https://arxiv.org/abs/2507.15820} {arXiv:2507.15820
  [hep-ph]} \BibitemShut {NoStop}%
\bibitem [{\citenamefont {Chang}\ \emph {et~al.}(2026)\citenamefont {Chang},
  \citenamefont {Chen}, \citenamefont {Liu}, \citenamefont {Simmons-Duffin},
  \citenamefont {Yuan},\ and\ \citenamefont {Zhu}}]{Chang:2025kgq}%
  \BibitemOpen
  \bibfield  {author} {\bibinfo {author} {\bibfnamefont {C.-H.}\ \bibnamefont
  {Chang}}, \bibinfo {author} {\bibfnamefont {H.}~\bibnamefont {Chen}},
  \bibinfo {author} {\bibfnamefont {X.}~\bibnamefont {Liu}}, \bibinfo {author}
  {\bibfnamefont {D.}~\bibnamefont {Simmons-Duffin}}, \bibinfo {author}
  {\bibfnamefont {F.}~\bibnamefont {Yuan}},\ and\ \bibinfo {author}
  {\bibfnamefont {H.~X.}\ \bibnamefont {Zhu}},\ }\bibfield  {title} {\bibinfo
  {title} {{Quantum Scaling in Energy Correlators beyond the Confinement
  Transition}},\ }\href {https://doi.org/10.1103/9ml8-xkfc} {\bibfield
  {journal} {\bibinfo  {journal} {Phys. Rev. Lett.}\ }\textbf {\bibinfo
  {volume} {136}},\ \bibinfo {pages} {081903} (\bibinfo {year} {2026})},\
  \Eprint {https://arxiv.org/abs/2507.15923} {arXiv:2507.15923 [hep-ph]}
  \BibitemShut {NoStop}%
\bibitem [{\citenamefont {Kang}\ \emph {et~al.}(2026)\citenamefont {Kang},
  \citenamefont {Metz}, \citenamefont {Pitonyak},\ and\ \citenamefont
  {Zhang}}]{Kang:2025zto}%
  \BibitemOpen
  \bibfield  {author} {\bibinfo {author} {\bibfnamefont {Z.-B.}\ \bibnamefont
  {Kang}}, \bibinfo {author} {\bibfnamefont {A.}~\bibnamefont {Metz}}, \bibinfo
  {author} {\bibfnamefont {D.}~\bibnamefont {Pitonyak}},\ and\ \bibinfo
  {author} {\bibfnamefont {C.}~\bibnamefont {Zhang}},\ }\bibfield  {title}
  {\bibinfo {title} {{Dihadron Fragmentation Framework for Near-Side
  Energy-Energy Correlators}},\ }\href {https://doi.org/10.1103/jnl4-x77t}
  {\bibfield  {journal} {\bibinfo  {journal} {Phys. Rev. Lett.}\ }\textbf
  {\bibinfo {volume} {136}},\ \bibinfo {pages} {081905} (\bibinfo {year}
  {2026})},\ \Eprint {https://arxiv.org/abs/2507.17444} {arXiv:2507.17444
  [hep-ph]} \BibitemShut {NoStop}%
\bibitem [{\citenamefont {Riembau}\ and\ \citenamefont
  {Son}(2026)}]{Riembau:2025wjc}%
  \BibitemOpen
  \bibfield  {author} {\bibinfo {author} {\bibfnamefont {M.}~\bibnamefont
  {Riembau}}\ and\ \bibinfo {author} {\bibfnamefont {M.}~\bibnamefont {Son}},\
  }\bibfield  {title} {\bibinfo {title} {{Flow between Extremal One-Point
  Energy Correlators in QCD}},\ }\href {https://doi.org/10.1103/lc5m-kwgv}
  {\bibfield  {journal} {\bibinfo  {journal} {Phys. Rev. Lett.}\ }\textbf
  {\bibinfo {volume} {136}},\ \bibinfo {pages} {101901} (\bibinfo {year}
  {2026})},\ \Eprint {https://arxiv.org/abs/2509.02669} {arXiv:2509.02669
  [hep-ph]} \BibitemShut {NoStop}%
\bibitem [{\citenamefont {Hofman}\ and\ \citenamefont
  {Maldacena}(2008)}]{Hofman:2008ar}%
  \BibitemOpen
  \bibfield  {author} {\bibinfo {author} {\bibfnamefont {D.~M.}\ \bibnamefont
  {Hofman}}\ and\ \bibinfo {author} {\bibfnamefont {J.}~\bibnamefont
  {Maldacena}},\ }\bibfield  {title} {\bibinfo {title} {{Conformal collider
  physics: Energy and charge correlations}},\ }\href
  {https://doi.org/10.1088/1126-6708/2008/05/012} {\bibfield  {journal}
  {\bibinfo  {journal} {JHEP}\ }\textbf {\bibinfo {volume} {05}},\ \bibinfo
  {pages} {012}},\ \Eprint {https://arxiv.org/abs/0803.1467} {arXiv:0803.1467
  [hep-th]} \BibitemShut {NoStop}%
\bibitem [{\citenamefont {Beisert}\ \emph {et~al.}(2007)\citenamefont
  {Beisert}, \citenamefont {Eden},\ and\ \citenamefont
  {Staudacher}}]{Beisert:2006ez}%
  \BibitemOpen
  \bibfield  {author} {\bibinfo {author} {\bibfnamefont {N.}~\bibnamefont
  {Beisert}}, \bibinfo {author} {\bibfnamefont {B.}~\bibnamefont {Eden}},\ and\
  \bibinfo {author} {\bibfnamefont {M.}~\bibnamefont {Staudacher}},\ }\bibfield
   {title} {\bibinfo {title} {{Transcendentality and Crossing}},\ }\href
  {https://doi.org/10.1088/1742-5468/2007/01/P01021} {\bibfield  {journal}
  {\bibinfo  {journal} {J. Stat. Mech.}\ }\textbf {\bibinfo {volume} {0701}},\
  \bibinfo {pages} {P01021} (\bibinfo {year} {2007})},\ \Eprint
  {https://arxiv.org/abs/hep-th/0610251} {arXiv:hep-th/0610251} \BibitemShut
  {NoStop}%
\bibitem [{\citenamefont {Basso}\ \emph {et~al.}(2008)\citenamefont {Basso},
  \citenamefont {Korchemsky},\ and\ \citenamefont {Kotanski}}]{Basso:2007wd}%
  \BibitemOpen
  \bibfield  {author} {\bibinfo {author} {\bibfnamefont {B.}~\bibnamefont
  {Basso}}, \bibinfo {author} {\bibfnamefont {G.~P.}\ \bibnamefont
  {Korchemsky}},\ and\ \bibinfo {author} {\bibfnamefont {J.}~\bibnamefont
  {Kotanski}},\ }\bibfield  {title} {\bibinfo {title} {{Cusp anomalous
  dimension in maximally supersymmetric Yang-Mills theory at strong
  coupling}},\ }\href {https://doi.org/10.1103/PhysRevLett.100.091601}
  {\bibfield  {journal} {\bibinfo  {journal} {Phys. Rev. Lett.}\ }\textbf
  {\bibinfo {volume} {100}},\ \bibinfo {pages} {091601} (\bibinfo {year}
  {2008})},\ \Eprint {https://arxiv.org/abs/0708.3933} {arXiv:0708.3933
  [hep-th]} \BibitemShut {NoStop}%
\bibitem [{\citenamefont {Gromov}\ \emph {et~al.}(2010)\citenamefont {Gromov},
  \citenamefont {Kazakov},\ and\ \citenamefont {Vieira}}]{Gromov:2009zb}%
  \BibitemOpen
  \bibfield  {author} {\bibinfo {author} {\bibfnamefont {N.}~\bibnamefont
  {Gromov}}, \bibinfo {author} {\bibfnamefont {V.}~\bibnamefont {Kazakov}},\
  and\ \bibinfo {author} {\bibfnamefont {P.}~\bibnamefont {Vieira}},\
  }\bibfield  {title} {\bibinfo {title} {{Exact Spectrum of Planar ${\cal N}=4$
  Supersymmetric Yang-Mills Theory: Konishi Dimension at Any Coupling}},\
  }\href {https://doi.org/10.1103/PhysRevLett.104.211601} {\bibfield  {journal}
  {\bibinfo  {journal} {Phys. Rev. Lett.}\ }\textbf {\bibinfo {volume} {104}},\
  \bibinfo {pages} {211601} (\bibinfo {year} {2010})},\ \Eprint
  {https://arxiv.org/abs/0906.4240} {arXiv:0906.4240 [hep-th]} \BibitemShut
  {NoStop}%
\bibitem [{\citenamefont {Gromov}\ \emph {et~al.}(2014)\citenamefont {Gromov},
  \citenamefont {Kazakov}, \citenamefont {Leurent},\ and\ \citenamefont
  {Volin}}]{Gromov:2013pga}%
  \BibitemOpen
  \bibfield  {author} {\bibinfo {author} {\bibfnamefont {N.}~\bibnamefont
  {Gromov}}, \bibinfo {author} {\bibfnamefont {V.}~\bibnamefont {Kazakov}},
  \bibinfo {author} {\bibfnamefont {S.}~\bibnamefont {Leurent}},\ and\ \bibinfo
  {author} {\bibfnamefont {D.}~\bibnamefont {Volin}},\ }\bibfield  {title}
  {\bibinfo {title} {{Quantum Spectral Curve for Planar $\mathcal{N} = 4$
  Super-Yang-Mills Theory}},\ }\href
  {https://doi.org/10.1103/PhysRevLett.112.011602} {\bibfield  {journal}
  {\bibinfo  {journal} {Phys. Rev. Lett.}\ }\textbf {\bibinfo {volume} {112}},\
  \bibinfo {pages} {011602} (\bibinfo {year} {2014})},\ \Eprint
  {https://arxiv.org/abs/1305.1939} {arXiv:1305.1939 [hep-th]} \BibitemShut
  {NoStop}%
\bibitem [{\citenamefont {Basso}\ \emph {et~al.}(2020)\citenamefont {Basso},
  \citenamefont {Dixon},\ and\ \citenamefont {Papathanasiou}}]{Basso:2020xts}%
  \BibitemOpen
  \bibfield  {author} {\bibinfo {author} {\bibfnamefont {B.}~\bibnamefont
  {Basso}}, \bibinfo {author} {\bibfnamefont {L.~J.}\ \bibnamefont {Dixon}},\
  and\ \bibinfo {author} {\bibfnamefont {G.}~\bibnamefont {Papathanasiou}},\
  }\bibfield  {title} {\bibinfo {title} {{Origin of the Six-Gluon Amplitude in
  Planar $N=4$ Supersymmetric Yang-Mills Theory}},\ }\href
  {https://doi.org/10.1103/PhysRevLett.124.161603} {\bibfield  {journal}
  {\bibinfo  {journal} {Phys. Rev. Lett.}\ }\textbf {\bibinfo {volume} {124}},\
  \bibinfo {pages} {161603} (\bibinfo {year} {2020})},\ \Eprint
  {https://arxiv.org/abs/2001.05460} {arXiv:2001.05460 [hep-th]} \BibitemShut
  {NoStop}%
\bibitem [{\citenamefont {Alday}\ \emph {et~al.}(2025)\citenamefont {Alday},
  \citenamefont {Armanini}, \citenamefont {H{\"a}ring},\ and\ \citenamefont
  {Zhiboedov}}]{Alday:2025pmg}%
  \BibitemOpen
  \bibfield  {author} {\bibinfo {author} {\bibfnamefont {L.~F.}\ \bibnamefont
  {Alday}}, \bibinfo {author} {\bibfnamefont {E.}~\bibnamefont {Armanini}},
  \bibinfo {author} {\bibfnamefont {K.}~\bibnamefont {H{\"a}ring}},\ and\
  \bibinfo {author} {\bibfnamefont {A.}~\bibnamefont {Zhiboedov}},\ }\bibfield
  {title} {\bibinfo {title} {{From Partons to Strings: Scattering on the
  Coulomb Branch of $\mathcal{N}=4$ SYM}},\ }\href@noop {} {\  (\bibinfo {year}
  {2025})},\ \Eprint {https://arxiv.org/abs/2510.19909} {arXiv:2510.19909
  [hep-th]} \BibitemShut {NoStop}%
\bibitem [{\citenamefont {Belitsky}\ \emph
  {et~al.}(2014{\natexlab{a}})\citenamefont {Belitsky}, \citenamefont
  {Hohenegger}, \citenamefont {Korchemsky}, \citenamefont {Sokatchev},\ and\
  \citenamefont {Zhiboedov}}]{Belitsky:2013xxa}%
  \BibitemOpen
  \bibfield  {author} {\bibinfo {author} {\bibfnamefont {A.~V.}\ \bibnamefont
  {Belitsky}}, \bibinfo {author} {\bibfnamefont {S.}~\bibnamefont
  {Hohenegger}}, \bibinfo {author} {\bibfnamefont {G.~P.}\ \bibnamefont
  {Korchemsky}}, \bibinfo {author} {\bibfnamefont {E.}~\bibnamefont
  {Sokatchev}},\ and\ \bibinfo {author} {\bibfnamefont {A.}~\bibnamefont
  {Zhiboedov}},\ }\bibfield  {title} {\bibinfo {title} {{From correlation
  functions to event shapes}},\ }\href
  {https://doi.org/10.1016/j.nuclphysb.2014.04.020} {\bibfield  {journal}
  {\bibinfo  {journal} {Nucl. Phys. B}\ }\textbf {\bibinfo {volume} {884}},\
  \bibinfo {pages} {305} (\bibinfo {year} {2014}{\natexlab{a}})},\ \Eprint
  {https://arxiv.org/abs/1309.0769} {arXiv:1309.0769 [hep-th]} \BibitemShut
  {NoStop}%
\bibitem [{\citenamefont {Belitsky}\ \emph
  {et~al.}(2014{\natexlab{b}})\citenamefont {Belitsky}, \citenamefont
  {Hohenegger}, \citenamefont {Korchemsky}, \citenamefont {Sokatchev},\ and\
  \citenamefont {Zhiboedov}}]{Belitsky:2013bja}%
  \BibitemOpen
  \bibfield  {author} {\bibinfo {author} {\bibfnamefont {A.~V.}\ \bibnamefont
  {Belitsky}}, \bibinfo {author} {\bibfnamefont {S.}~\bibnamefont
  {Hohenegger}}, \bibinfo {author} {\bibfnamefont {G.~P.}\ \bibnamefont
  {Korchemsky}}, \bibinfo {author} {\bibfnamefont {E.}~\bibnamefont
  {Sokatchev}},\ and\ \bibinfo {author} {\bibfnamefont {A.}~\bibnamefont
  {Zhiboedov}},\ }\bibfield  {title} {\bibinfo {title} {{Event shapes in
  $\mathcal{N} = 4$ super-Yang-Mills theory}},\ }\href
  {https://doi.org/10.1016/j.nuclphysb.2014.04.019} {\bibfield  {journal}
  {\bibinfo  {journal} {Nucl. Phys. B}\ }\textbf {\bibinfo {volume} {884}},\
  \bibinfo {pages} {206} (\bibinfo {year} {2014}{\natexlab{b}})},\ \Eprint
  {https://arxiv.org/abs/1309.1424} {arXiv:1309.1424 [hep-th]} \BibitemShut
  {NoStop}%
\bibitem [{\citenamefont {Belitsky}\ \emph
  {et~al.}(2014{\natexlab{c}})\citenamefont {Belitsky}, \citenamefont
  {Hohenegger}, \citenamefont {Korchemsky}, \citenamefont {Sokatchev},\ and\
  \citenamefont {Zhiboedov}}]{Belitsky:2013ofa}%
  \BibitemOpen
  \bibfield  {author} {\bibinfo {author} {\bibfnamefont {A.~V.}\ \bibnamefont
  {Belitsky}}, \bibinfo {author} {\bibfnamefont {S.}~\bibnamefont
  {Hohenegger}}, \bibinfo {author} {\bibfnamefont {G.~P.}\ \bibnamefont
  {Korchemsky}}, \bibinfo {author} {\bibfnamefont {E.}~\bibnamefont
  {Sokatchev}},\ and\ \bibinfo {author} {\bibfnamefont {A.}~\bibnamefont
  {Zhiboedov}},\ }\bibfield  {title} {\bibinfo {title} {{Energy-Energy
  Correlations in N=4 Supersymmetric Yang-Mills Theory}},\ }\href
  {https://doi.org/10.1103/PhysRevLett.112.071601} {\bibfield  {journal}
  {\bibinfo  {journal} {Phys. Rev. Lett.}\ }\textbf {\bibinfo {volume} {112}},\
  \bibinfo {pages} {071601} (\bibinfo {year} {2014}{\natexlab{c}})},\ \Eprint
  {https://arxiv.org/abs/1311.6800} {arXiv:1311.6800 [hep-th]} \BibitemShut
  {NoStop}%
\bibitem [{\citenamefont {Dempsey}\ \emph {et~al.}(2025)\citenamefont
  {Dempsey}, \citenamefont {Karlsson}, \citenamefont {Pufu}, \citenamefont
  {Zahraee},\ and\ \citenamefont {Zhiboedov}}]{Dempsey:2025yiv}%
  \BibitemOpen
  \bibfield  {author} {\bibinfo {author} {\bibfnamefont {R.}~\bibnamefont
  {Dempsey}}, \bibinfo {author} {\bibfnamefont {R.}~\bibnamefont {Karlsson}},
  \bibinfo {author} {\bibfnamefont {S.~S.}\ \bibnamefont {Pufu}}, \bibinfo
  {author} {\bibfnamefont {Z.}~\bibnamefont {Zahraee}},\ and\ \bibinfo {author}
  {\bibfnamefont {A.}~\bibnamefont {Zhiboedov}},\ }\bibfield  {title} {\bibinfo
  {title} {{Conformal collider bootstrap in ${\mathcal N}=4$ SYM}},\
  }\href@noop {} {\  (\bibinfo {year} {2025})},\ \Eprint
  {https://arxiv.org/abs/2512.10796} {arXiv:2512.10796 [hep-th]} \BibitemShut
  {NoStop}%
\bibitem [{\citenamefont {Alday}\ and\ \citenamefont
  {Maldacena}(2007)}]{Alday:2007mf}%
  \BibitemOpen
  \bibfield  {author} {\bibinfo {author} {\bibfnamefont {L.~F.}\ \bibnamefont
  {Alday}}\ and\ \bibinfo {author} {\bibfnamefont {J.~M.}\ \bibnamefont
  {Maldacena}},\ }\bibfield  {title} {\bibinfo {title} {{Comments on operators
  with large spin}},\ }\href {https://doi.org/10.1088/1126-6708/2007/11/019}
  {\bibfield  {journal} {\bibinfo  {journal} {JHEP}\ }\textbf {\bibinfo
  {volume} {11}},\ \bibinfo {pages} {019}},\ \Eprint
  {https://arxiv.org/abs/0708.0672} {arXiv:0708.0672 [hep-th]} \BibitemShut
  {NoStop}%
\bibitem [{\citenamefont {Gubser}\ \emph {et~al.}(2002)\citenamefont {Gubser},
  \citenamefont {Klebanov},\ and\ \citenamefont {Polyakov}}]{Gubser:2002tv}%
  \BibitemOpen
  \bibfield  {author} {\bibinfo {author} {\bibfnamefont {S.~S.}\ \bibnamefont
  {Gubser}}, \bibinfo {author} {\bibfnamefont {I.~R.}\ \bibnamefont
  {Klebanov}},\ and\ \bibinfo {author} {\bibfnamefont {A.~M.}\ \bibnamefont
  {Polyakov}},\ }\bibfield  {title} {\bibinfo {title} {{A Semiclassical limit
  of the gauge / string correspondence}},\ }\href
  {https://doi.org/10.1016/S0550-3213(02)00373-5} {\bibfield  {journal}
  {\bibinfo  {journal} {Nucl. Phys. B}\ }\textbf {\bibinfo {volume} {636}},\
  \bibinfo {pages} {99} (\bibinfo {year} {2002})},\ \Eprint
  {https://arxiv.org/abs/hep-th/0204051} {arXiv:hep-th/0204051} \BibitemShut
  {NoStop}%
\bibitem [{\citenamefont {Klebanov}\ \emph {et~al.}(2006)\citenamefont
  {Klebanov}, \citenamefont {Maldacena},\ and\ \citenamefont
  {Thorn}}]{Klebanov:2006jj}%
  \BibitemOpen
  \bibfield  {author} {\bibinfo {author} {\bibfnamefont {I.~R.}\ \bibnamefont
  {Klebanov}}, \bibinfo {author} {\bibfnamefont {J.~M.}\ \bibnamefont
  {Maldacena}},\ and\ \bibinfo {author} {\bibfnamefont {C.~B.}\ \bibnamefont
  {Thorn}, \bibfnamefont {III}},\ }\bibfield  {title} {\bibinfo {title}
  {{Dynamics of flux tubes in large N gauge theories}},\ }\href
  {https://doi.org/10.1088/1126-6708/2006/04/024} {\bibfield  {journal}
  {\bibinfo  {journal} {JHEP}\ }\textbf {\bibinfo {volume} {04}},\ \bibinfo
  {pages} {024}},\ \Eprint {https://arxiv.org/abs/hep-th/0602255}
  {arXiv:hep-th/0602255} \BibitemShut {NoStop}%
\bibitem [{\citenamefont {Korchemsky}(2020)}]{Korchemsky:2019nzm}%
  \BibitemOpen
  \bibfield  {author} {\bibinfo {author} {\bibfnamefont {G.~P.}\ \bibnamefont
  {Korchemsky}},\ }\bibfield  {title} {\bibinfo {title} {{Energy correlations
  in the end-point region}},\ }\href {https://doi.org/10.1007/JHEP01(2020)008}
  {\bibfield  {journal} {\bibinfo  {journal} {JHEP}\ }\textbf {\bibinfo
  {volume} {01}},\ \bibinfo {pages} {008}},\ \Eprint
  {https://arxiv.org/abs/1905.01444} {arXiv:1905.01444 [hep-th]} \BibitemShut
  {NoStop}%
\bibitem [{\citenamefont {Dokshitzer}\ \emph {et~al.}(1999)\citenamefont
  {Dokshitzer}, \citenamefont {Marchesini},\ and\ \citenamefont
  {Webber}}]{Dokshitzer:1999sh}%
  \BibitemOpen
  \bibfield  {author} {\bibinfo {author} {\bibfnamefont {Y.~L.}\ \bibnamefont
  {Dokshitzer}}, \bibinfo {author} {\bibfnamefont {G.}~\bibnamefont
  {Marchesini}},\ and\ \bibinfo {author} {\bibfnamefont {B.~R.}\ \bibnamefont
  {Webber}},\ }\bibfield  {title} {\bibinfo {title} {{Nonperturbative effects
  in the energy energy correlation}},\ }\href
  {https://doi.org/10.1088/1126-6708/1999/07/012} {\bibfield  {journal}
  {\bibinfo  {journal} {JHEP}\ }\textbf {\bibinfo {volume} {07}},\ \bibinfo
  {pages} {012}},\ \Eprint {https://arxiv.org/abs/hep-ph/9905339}
  {arXiv:hep-ph/9905339} \BibitemShut {NoStop}%
\bibitem [{\citenamefont {Alday}\ \emph {et~al.}(2015)\citenamefont {Alday},
  \citenamefont {Bissi},\ and\ \citenamefont {Lukowski}}]{Alday:2015eya}%
  \BibitemOpen
  \bibfield  {author} {\bibinfo {author} {\bibfnamefont {L.~F.}\ \bibnamefont
  {Alday}}, \bibinfo {author} {\bibfnamefont {A.}~\bibnamefont {Bissi}},\ and\
  \bibinfo {author} {\bibfnamefont {T.}~\bibnamefont {Lukowski}},\ }\bibfield
  {title} {\bibinfo {title} {{Large spin systematics in CFT}},\ }\href
  {https://doi.org/10.1007/JHEP11(2015)101} {\bibfield  {journal} {\bibinfo
  {journal} {JHEP}\ }\textbf {\bibinfo {volume} {11}},\ \bibinfo {pages}
  {101}},\ \Eprint {https://arxiv.org/abs/1502.07707} {arXiv:1502.07707
  [hep-th]} \BibitemShut {NoStop}%
\bibitem [{\citenamefont {Alday}(2017)}]{Alday:2016njk}%
  \BibitemOpen
  \bibfield  {author} {\bibinfo {author} {\bibfnamefont {L.~F.}\ \bibnamefont
  {Alday}},\ }\bibfield  {title} {\bibinfo {title} {{Large Spin Perturbation
  Theory for Conformal Field Theories}},\ }\href
  {https://doi.org/10.1103/PhysRevLett.119.111601} {\bibfield  {journal}
  {\bibinfo  {journal} {Phys. Rev. Lett.}\ }\textbf {\bibinfo {volume} {119}},\
  \bibinfo {pages} {111601} (\bibinfo {year} {2017})},\ \Eprint
  {https://arxiv.org/abs/1611.01500} {arXiv:1611.01500 [hep-th]} \BibitemShut
  {NoStop}%
\bibitem [{\citenamefont {Henriksson}\ and\ \citenamefont
  {Lukowski}(2018)}]{Henriksson:2017eej}%
  \BibitemOpen
  \bibfield  {author} {\bibinfo {author} {\bibfnamefont {J.}~\bibnamefont
  {Henriksson}}\ and\ \bibinfo {author} {\bibfnamefont {T.}~\bibnamefont
  {Lukowski}},\ }\bibfield  {title} {\bibinfo {title} {{Perturbative Four-Point
  Functions from the Analytic Conformal Bootstrap}},\ }\href
  {https://doi.org/10.1007/JHEP02(2018)123} {\bibfield  {journal} {\bibinfo
  {journal} {JHEP}\ }\textbf {\bibinfo {volume} {02}},\ \bibinfo {pages}
  {123}},\ \Eprint {https://arxiv.org/abs/1710.06242} {arXiv:1710.06242
  [hep-th]} \BibitemShut {NoStop}%
\bibitem [{\citenamefont {Chen}\ \emph {et~al.}(2023)\citenamefont {Chen},
  \citenamefont {Zhou},\ and\ \citenamefont {Zhu}}]{Chen:2023wah}%
  \BibitemOpen
  \bibfield  {author} {\bibinfo {author} {\bibfnamefont {H.}~\bibnamefont
  {Chen}}, \bibinfo {author} {\bibfnamefont {X.}~\bibnamefont {Zhou}},\ and\
  \bibinfo {author} {\bibfnamefont {H.~X.}\ \bibnamefont {Zhu}},\ }\bibfield
  {title} {\bibinfo {title} {{Power corrections to energy flow correlations
  from large spin perturbation}},\ }\href
  {https://doi.org/10.1007/JHEP10(2023)132} {\bibfield  {journal} {\bibinfo
  {journal} {JHEP}\ }\textbf {\bibinfo {volume} {10}},\ \bibinfo {pages}
  {132}},\ \Eprint {https://arxiv.org/abs/2301.03616} {arXiv:2301.03616
  [hep-ph]} \BibitemShut {NoStop}%
\bibitem [{\citenamefont {Chen}\ \emph {et~al.}(2024)\citenamefont {Chen},
  \citenamefont {Zhou},\ and\ \citenamefont {Zhu}}]{Chen:2023amz}%
  \BibitemOpen
  \bibfield  {author} {\bibinfo {author} {\bibfnamefont {H.}~\bibnamefont
  {Chen}}, \bibinfo {author} {\bibfnamefont {X.}~\bibnamefont {Zhou}},\ and\
  \bibinfo {author} {\bibfnamefont {H.~X.}\ \bibnamefont {Zhu}},\ }\bibfield
  {title} {\bibinfo {title} {{Power corrections to EEC meets conformal
  bootstrap}},\ }\href {https://doi.org/10.22323/1.432.0061} {\bibfield
  {journal} {\bibinfo  {journal} {PoS}\ }\textbf {\bibinfo {volume}
  {RADCOR2023}},\ \bibinfo {pages} {061} (\bibinfo {year} {2024})}\BibitemShut
  {NoStop}%
\bibitem [{\citenamefont {Korchemsky}\ \emph {et~al.}(1997)\citenamefont
  {Korchemsky}, \citenamefont {Oderda},\ and\ \citenamefont
  {Sterman}}]{Korchemsky:1997sy}%
  \BibitemOpen
  \bibfield  {author} {\bibinfo {author} {\bibfnamefont {G.~P.}\ \bibnamefont
  {Korchemsky}}, \bibinfo {author} {\bibfnamefont {G.}~\bibnamefont {Oderda}},\
  and\ \bibinfo {author} {\bibfnamefont {G.~F.}\ \bibnamefont {Sterman}},\
  }\bibfield  {title} {\bibinfo {title} {{Power corrections and nonlocal
  operators}},\ }\href {https://doi.org/10.1063/1.53732} {\bibfield  {journal}
  {\bibinfo  {journal} {AIP Conf. Proc.}\ }\textbf {\bibinfo {volume} {407}},\
  \bibinfo {pages} {988} (\bibinfo {year} {1997})},\ \Eprint
  {https://arxiv.org/abs/hep-ph/9708346} {arXiv:hep-ph/9708346} \BibitemShut
  {NoStop}%
\bibitem [{\citenamefont {Sveshnikov}\ and\ \citenamefont
  {Tkachov}(1996)}]{Sveshnikov:1995vi}%
  \BibitemOpen
  \bibfield  {author} {\bibinfo {author} {\bibfnamefont {N.~A.}\ \bibnamefont
  {Sveshnikov}}\ and\ \bibinfo {author} {\bibfnamefont {F.~V.}\ \bibnamefont
  {Tkachov}},\ }\bibfield  {title} {\bibinfo {title} {{Jets and quantum field
  theory}},\ }\href {https://doi.org/10.1016/0370-2693(96)00558-8} {\bibfield
  {journal} {\bibinfo  {journal} {Phys. Lett. B}\ }\textbf {\bibinfo {volume}
  {382}},\ \bibinfo {pages} {403} (\bibinfo {year} {1996})},\ \Eprint
  {https://arxiv.org/abs/hep-ph/9512370} {arXiv:hep-ph/9512370} \BibitemShut
  {NoStop}%
\bibitem [{\citenamefont {Kotikov}\ \emph {et~al.}(2004)\citenamefont
  {Kotikov}, \citenamefont {Lipatov}, \citenamefont {Onishchenko},\ and\
  \citenamefont {Velizhanin}}]{Kotikov:2004er}%
  \BibitemOpen
  \bibfield  {author} {\bibinfo {author} {\bibfnamefont {A.~V.}\ \bibnamefont
  {Kotikov}}, \bibinfo {author} {\bibfnamefont {L.~N.}\ \bibnamefont
  {Lipatov}}, \bibinfo {author} {\bibfnamefont {A.~I.}\ \bibnamefont
  {Onishchenko}},\ and\ \bibinfo {author} {\bibfnamefont {V.~N.}\ \bibnamefont
  {Velizhanin}},\ }\bibfield  {title} {\bibinfo {title} {{Three loop universal
  anomalous dimension of the Wilson operators in $N=4$ SUSY Yang-Mills
  model}},\ }\href {https://doi.org/10.1016/j.physletb.2004.05.078} {\bibfield
  {journal} {\bibinfo  {journal} {Phys. Lett. B}\ }\textbf {\bibinfo {volume}
  {595}},\ \bibinfo {pages} {521} (\bibinfo {year} {2004})},\ \bibinfo {note}
  {[Erratum: Phys.Lett.B 632, 754--756 (2006)]},\ \Eprint
  {https://arxiv.org/abs/hep-th/0404092} {arXiv:hep-th/0404092} \BibitemShut
  {NoStop}%
\bibitem [{\citenamefont {Freyhult}\ \emph {et~al.}(2008)\citenamefont
  {Freyhult}, \citenamefont {Rej},\ and\ \citenamefont
  {Staudacher}}]{Freyhult:2007pz}%
  \BibitemOpen
  \bibfield  {author} {\bibinfo {author} {\bibfnamefont {L.}~\bibnamefont
  {Freyhult}}, \bibinfo {author} {\bibfnamefont {A.}~\bibnamefont {Rej}},\ and\
  \bibinfo {author} {\bibfnamefont {M.}~\bibnamefont {Staudacher}},\ }\bibfield
   {title} {\bibinfo {title} {{A Generalized Scaling Function for AdS/CFT}},\
  }\href {https://doi.org/10.1088/1742-5468/2008/07/P07015} {\bibfield
  {journal} {\bibinfo  {journal} {J. Stat. Mech.}\ }\textbf {\bibinfo {volume}
  {0807}},\ \bibinfo {pages} {P07015} (\bibinfo {year} {2008})},\ \Eprint
  {https://arxiv.org/abs/0712.2743} {arXiv:0712.2743 [hep-th]} \BibitemShut
  {NoStop}%
\bibitem [{\citenamefont {Grozin}\ \emph {et~al.}(2016)\citenamefont {Grozin},
  \citenamefont {Henn}, \citenamefont {Korchemsky},\ and\ \citenamefont
  {Marquard}}]{Grozin:2015kna}%
  \BibitemOpen
  \bibfield  {author} {\bibinfo {author} {\bibfnamefont {A.}~\bibnamefont
  {Grozin}}, \bibinfo {author} {\bibfnamefont {J.~M.}\ \bibnamefont {Henn}},
  \bibinfo {author} {\bibfnamefont {G.~P.}\ \bibnamefont {Korchemsky}},\ and\
  \bibinfo {author} {\bibfnamefont {P.}~\bibnamefont {Marquard}},\ }\bibfield
  {title} {\bibinfo {title} {{The three-loop cusp anomalous dimension in QCD
  and its supersymmetric extensions}},\ }\href
  {https://doi.org/10.1007/JHEP01(2016)140} {\bibfield  {journal} {\bibinfo
  {journal} {JHEP}\ }\textbf {\bibinfo {volume} {01}},\ \bibinfo {pages}
  {140}},\ \Eprint {https://arxiv.org/abs/1510.07803} {arXiv:1510.07803
  [hep-ph]} \BibitemShut {NoStop}%
\bibitem [{\citenamefont {Henn}\ \emph {et~al.}(2020)\citenamefont {Henn},
  \citenamefont {Korchemsky},\ and\ \citenamefont
  {Mistlberger}}]{Henn:2019swt}%
  \BibitemOpen
  \bibfield  {author} {\bibinfo {author} {\bibfnamefont {J.~M.}\ \bibnamefont
  {Henn}}, \bibinfo {author} {\bibfnamefont {G.~P.}\ \bibnamefont
  {Korchemsky}},\ and\ \bibinfo {author} {\bibfnamefont {B.}~\bibnamefont
  {Mistlberger}},\ }\bibfield  {title} {\bibinfo {title} {{The full four-loop
  cusp anomalous dimension in $\mathcal{N}=4$ super Yang-Mills and QCD}},\
  }\href {https://doi.org/10.1007/JHEP04(2020)018} {\bibfield  {journal}
  {\bibinfo  {journal} {JHEP}\ }\textbf {\bibinfo {volume} {04}},\ \bibinfo
  {pages} {018}},\ \Eprint {https://arxiv.org/abs/1911.10174} {arXiv:1911.10174
  [hep-th]} \BibitemShut {NoStop}%
\bibitem [{\citenamefont {Freyhult}\ and\ \citenamefont
  {Zieme}(2009)}]{Freyhult:2009my}%
  \BibitemOpen
  \bibfield  {author} {\bibinfo {author} {\bibfnamefont {L.}~\bibnamefont
  {Freyhult}}\ and\ \bibinfo {author} {\bibfnamefont {S.}~\bibnamefont
  {Zieme}},\ }\bibfield  {title} {\bibinfo {title} {{The virtual scaling
  function of AdS/CFT}},\ }\href {https://doi.org/10.1103/PhysRevD.79.105009}
  {\bibfield  {journal} {\bibinfo  {journal} {Phys. Rev. D}\ }\textbf {\bibinfo
  {volume} {79}},\ \bibinfo {pages} {105009} (\bibinfo {year} {2009})},\
  \Eprint {https://arxiv.org/abs/0901.2749} {arXiv:0901.2749 [hep-th]}
  \BibitemShut {NoStop}%
\bibitem [{\citenamefont {Eden}(2012)}]{Eden:2012rr}%
  \BibitemOpen
  \bibfield  {author} {\bibinfo {author} {\bibfnamefont {B.}~\bibnamefont
  {Eden}},\ }\bibfield  {title} {\bibinfo {title} {{Three-loop universal
  structure constants in N=4 susy Yang-Mills theory}},\ }\href@noop {} {\
  (\bibinfo {year} {2012})},\ \Eprint {https://arxiv.org/abs/1207.3112}
  {arXiv:1207.3112 [hep-th]} \BibitemShut {NoStop}%
\bibitem [{\citenamefont {Alday}\ and\ \citenamefont
  {Bissi}(2013)}]{Alday:2013cwa}%
  \BibitemOpen
  \bibfield  {author} {\bibinfo {author} {\bibfnamefont {L.~F.}\ \bibnamefont
  {Alday}}\ and\ \bibinfo {author} {\bibfnamefont {A.}~\bibnamefont {Bissi}},\
  }\bibfield  {title} {\bibinfo {title} {{Higher-spin correlators}},\ }\href
  {https://doi.org/10.1007/JHEP10(2013)202} {\bibfield  {journal} {\bibinfo
  {journal} {JHEP}\ }\textbf {\bibinfo {volume} {10}},\ \bibinfo {pages}
  {202}},\ \Eprint {https://arxiv.org/abs/1305.4604} {arXiv:1305.4604 [hep-th]}
  \BibitemShut {NoStop}%
\bibitem [{\citenamefont {Kotikov}\ \emph {et~al.}(2007)\citenamefont
  {Kotikov}, \citenamefont {Lipatov}, \citenamefont {Rej}, \citenamefont
  {Staudacher},\ and\ \citenamefont {Velizhanin}}]{Kotikov:2007cy}%
  \BibitemOpen
  \bibfield  {author} {\bibinfo {author} {\bibfnamefont {A.~V.}\ \bibnamefont
  {Kotikov}}, \bibinfo {author} {\bibfnamefont {L.~N.}\ \bibnamefont
  {Lipatov}}, \bibinfo {author} {\bibfnamefont {A.}~\bibnamefont {Rej}},
  \bibinfo {author} {\bibfnamefont {M.}~\bibnamefont {Staudacher}},\ and\
  \bibinfo {author} {\bibfnamefont {V.~N.}\ \bibnamefont {Velizhanin}},\
  }\bibfield  {title} {\bibinfo {title} {{Dressing and wrapping}},\ }\href
  {https://doi.org/10.1088/1742-5468/2007/10/P10003} {\bibfield  {journal}
  {\bibinfo  {journal} {J. Stat. Mech.}\ }\textbf {\bibinfo {volume} {0710}},\
  \bibinfo {pages} {P10003} (\bibinfo {year} {2007})},\ \Eprint
  {https://arxiv.org/abs/0704.3586} {arXiv:0704.3586 [hep-th]} \BibitemShut
  {NoStop}%
\bibitem [{\citenamefont {Bajnok}\ \emph {et~al.}(2009)\citenamefont {Bajnok},
  \citenamefont {Janik},\ and\ \citenamefont {Lukowski}}]{Bajnok:2008qj}%
  \BibitemOpen
  \bibfield  {author} {\bibinfo {author} {\bibfnamefont {Z.}~\bibnamefont
  {Bajnok}}, \bibinfo {author} {\bibfnamefont {R.~A.}\ \bibnamefont {Janik}},\
  and\ \bibinfo {author} {\bibfnamefont {T.}~\bibnamefont {Lukowski}},\
  }\bibfield  {title} {\bibinfo {title} {{Four loop twist two, BFKL, wrapping
  and strings}},\ }\href {https://doi.org/10.1016/j.nuclphysb.2009.02.005}
  {\bibfield  {journal} {\bibinfo  {journal} {Nucl. Phys. B}\ }\textbf
  {\bibinfo {volume} {816}},\ \bibinfo {pages} {376} (\bibinfo {year}
  {2009})},\ \Eprint {https://arxiv.org/abs/0811.4448} {arXiv:0811.4448
  [hep-th]} \BibitemShut {NoStop}%
\bibitem [{\citenamefont {Beccaria}\ and\ \citenamefont
  {Forini}(2009)}]{Beccaria:2009vt}%
  \BibitemOpen
  \bibfield  {author} {\bibinfo {author} {\bibfnamefont {M.}~\bibnamefont
  {Beccaria}}\ and\ \bibinfo {author} {\bibfnamefont {V.}~\bibnamefont
  {Forini}},\ }\bibfield  {title} {\bibinfo {title} {{Four loop reciprocity of
  twist two operators in N=4 SYM}},\ }\href
  {https://doi.org/10.1088/1126-6708/2009/03/111} {\bibfield  {journal}
  {\bibinfo  {journal} {JHEP}\ }\textbf {\bibinfo {volume} {03}},\ \bibinfo
  {pages} {111}},\ \Eprint {https://arxiv.org/abs/0901.1256} {arXiv:0901.1256
  [hep-th]} \BibitemShut {NoStop}%
\bibitem [{\citenamefont {Kniehl}\ and\ \citenamefont
  {Velizhanin}(2021)}]{Kniehl:2021ysp}%
  \BibitemOpen
  \bibfield  {author} {\bibinfo {author} {\bibfnamefont {B.~A.}\ \bibnamefont
  {Kniehl}}\ and\ \bibinfo {author} {\bibfnamefont {V.~N.}\ \bibnamefont
  {Velizhanin}},\ }\bibfield  {title} {\bibinfo {title} {{Non-planar universal
  anomalous dimension of twist-two operators with general Lorentz spin at four
  loops in $N=4$ SYM theory}},\ }\href
  {https://doi.org/10.1016/j.nuclphysb.2021.115429} {\bibfield  {journal}
  {\bibinfo  {journal} {Nucl. Phys. B}\ }\textbf {\bibinfo {volume} {968}},\
  \bibinfo {pages} {115429} (\bibinfo {year} {2021})},\ \Eprint
  {https://arxiv.org/abs/2103.16420} {arXiv:2103.16420 [hep-th]} \BibitemShut
  {NoStop}%
\bibitem [{\citenamefont {Kniehl}\ and\ \citenamefont
  {Velizhanin}(2025)}]{Kniehl:2024tvd}%
  \BibitemOpen
  \bibfield  {author} {\bibinfo {author} {\bibfnamefont {B.~A.}\ \bibnamefont
  {Kniehl}}\ and\ \bibinfo {author} {\bibfnamefont {V.~N.}\ \bibnamefont
  {Velizhanin}},\ }\bibfield  {title} {\bibinfo {title} {{Nonplanar four-loop
  anomalous dimensions of twist-two operators in N=4 supersymmetric Yang-Mills
  theory: Higher moment, general result, and cusp anomalous dimension}},\
  }\href {https://doi.org/10.1103/PhysRevD.111.L061902} {\bibfield  {journal}
  {\bibinfo  {journal} {Phys. Rev. D}\ }\textbf {\bibinfo {volume} {111}},\
  \bibinfo {pages} {L061902} (\bibinfo {year} {2025})},\ \Eprint
  {https://arxiv.org/abs/2409.09463} {arXiv:2409.09463 [hep-th]} \BibitemShut
  {NoStop}%
\bibitem [{\citenamefont {Henn}\ \emph {et~al.}(2019)\citenamefont {Henn},
  \citenamefont {Sokatchev}, \citenamefont {Yan},\ and\ \citenamefont
  {Zhiboedov}}]{Henn:2019gkr}%
  \BibitemOpen
  \bibfield  {author} {\bibinfo {author} {\bibfnamefont {J.~M.}\ \bibnamefont
  {Henn}}, \bibinfo {author} {\bibfnamefont {E.}~\bibnamefont {Sokatchev}},
  \bibinfo {author} {\bibfnamefont {K.}~\bibnamefont {Yan}},\ and\ \bibinfo
  {author} {\bibfnamefont {A.}~\bibnamefont {Zhiboedov}},\ }\bibfield  {title}
  {\bibinfo {title} {{Energy-energy correlation in $N$=4 super Yang-Mills
  theory at next-to-next-to-leading order}},\ }\href
  {https://doi.org/10.1103/PhysRevD.100.036010} {\bibfield  {journal} {\bibinfo
   {journal} {Phys. Rev. D}\ }\textbf {\bibinfo {volume} {100}},\ \bibinfo
  {pages} {036010} (\bibinfo {year} {2019})},\ \Eprint
  {https://arxiv.org/abs/1903.05314} {arXiv:1903.05314 [hep-th]} \BibitemShut
  {NoStop}%
\bibitem [{\citenamefont {Ren}\ \emph {et~al.}(2026)\citenamefont {Ren},
  \citenamefont {Wang},\ and\ \citenamefont {Wen}}]{Ren:2026zxs}%
  \BibitemOpen
  \bibfield  {author} {\bibinfo {author} {\bibfnamefont {L.}~\bibnamefont
  {Ren}}, \bibinfo {author} {\bibfnamefont {B.}~\bibnamefont {Wang}},\ and\
  \bibinfo {author} {\bibfnamefont {C.}~\bibnamefont {Wen}},\ }\bibfield
  {title} {\bibinfo {title} {{Energy-energy correlator from the AdS
  Virasoro-Shapiro amplitude}},\ }\href {https://doi.org/10.1103/m9q6-47z2}
  {\bibfield  {journal} {\bibinfo  {journal} {Phys. Rev. D}\ }\textbf {\bibinfo
  {volume} {113}},\ \bibinfo {pages} {106013} (\bibinfo {year} {2026})},\
  \Eprint {https://arxiv.org/abs/2601.05312} {arXiv:2601.05312 [hep-th]}
  \BibitemShut {NoStop}%
\bibitem [{\citenamefont {Jackson}\ \emph {et~al.}(2026)\citenamefont
  {Jackson}, \citenamefont {Ren}, \citenamefont {Wang},\ and\ \citenamefont
  {Wen}}]{Jackson:2026eml}%
  \BibitemOpen
  \bibfield  {author} {\bibinfo {author} {\bibfnamefont {M.}~\bibnamefont
  {Jackson}}, \bibinfo {author} {\bibfnamefont {L.}~\bibnamefont {Ren}},
  \bibinfo {author} {\bibfnamefont {B.}~\bibnamefont {Wang}},\ and\ \bibinfo
  {author} {\bibfnamefont {C.}~\bibnamefont {Wen}},\ }\bibfield  {title}
  {\bibinfo {title} {{Energy-Energy Correlators at Strong Coupling}},\
  }\href@noop {} {\  (\bibinfo {year} {2026})},\ \Eprint
  {https://arxiv.org/abs/2607.21139} {arXiv:2607.21139 [hep-th]} \BibitemShut
  {NoStop}%
\end{thebibliography}%
    
    \widetext
	\begin{center}
		\textbf{\large Supplemental Materials}
	\end{center}
	\appendix

\section{Details of the N$^3$LL resummation}
This appendix gives the derivation logic behind the back-to-back endpoint coefficients used in the main text. We derive the LP and NLP logarithmic coefficients used in the main text, following the large-spin method of Refs.~\cite{Chen:2023wah,Chen:2023amz}.
The generic coefficients follow from twist-two data, while fixed-order endpoint expansions determine the exceptional low-order terms.

The EEC angular variable is $z=\sin^2(\theta/2)$, so the back-to-back limit is described by $y=1-z \to 0$, where we define $L_y=\log y$. Before performing light transform to obtain EEC, we use the standard conformal cross ratios $(z,\bar z)$, with $u=z\bar z$ and $v=(1-z)(1-\bar z)$ to describe local correlator. We write $L_u=\log u$ and $L_v=\log v$. In the crossed light-cone limit we set $\epsilon=1-\bar z$ and $L_\epsilon=\log\epsilon$. 

The all-order perturbative expansion of EEC in the back-to-back limit is organized as
\begin{align}
{\rm EEC}_{\rm b2b}(y)=\sum_{n=0}^\infty a^n {\rm EEC}^{(n)}_{\rm b2b}(y)
\,,\qquad
{\rm EEC}^{(n)}_{\rm b2b}(y)
&=\frac{1}{y}\sum_{k=1}^{2n}C^{\rm LP}_{k}(n)L_y^{2n-k}
+\sum_{k=1}^{2n}C^{\rm NLP}_{k}(n)L_y^{2n-k}
+\cdots .
\label{eq:b2b-coeff-map}
\end{align}
where the integer $n$ labels the perturbative order $a^n$, while $k=1,2,3,4$ labels the LL, NLL, NNLL and N$^3$LL in the back-to-back logarithmic tower. LP refers to the $y^{-1}$ sector and NLP to the $y^0$ sector. In the regime where $a\ll1$, $ |L_y|\gg1$, $aL_y^2 \sim\mathcal{O}(1)$, we re-organize to the resummation representation
\begin{align}
    {\rm EEC}_{\rm b2b}(y)=\sum_{k=1}^\infty
    \left[\frac{1}{y}\sum_{n\geq k/2}^\infty  a^n C^{\rm LP}_{k}(n)L_y^{2n-k}
    + \sum_{n\geq k/2}^\infty  a^n C^{\rm NLP}_{k}(n)L_y^{2n-k} +\cdots
    \right]
\end{align}
and in this work we truncate the sum of $k$ to be $k\leq 4$. 

\medskip
\noindent\emph{Block decomposition and twist-two data.} The dynamical part of the scalar correlator is expanded in superconformal blocks as
\begin{align}
\mathcal F(u,v)
=\sum_{\tau_0,\ell\ {\rm even}}
a_{\tau,\ell}\, G_{\tau+4,\ell}(z,\bar z),
\qquad
\mathcal F(u,v)-\mathcal F^{(0)}(u,v)
=\frac{u^3}{v}\Phi(u,v).
\label{eq:appendix-block-source}
\end{align}
Here $G_{\tau,\ell}$ is the normal $4d$ conformal block and $\mathcal{F}^{(0)}$ is the free correlator. The function $\Phi$ is crossing symmetric: $\Phi(u,v)=\Phi(v,u)$. As we show below, the back-to-back logarithms at LP and NLP are fully controlled by the twist-2 large-spin tail due to crossing symmetry.  We use the following perturbative ansatz for the twist-2 anomalous dimension~\cite{Kotikov:2004er,Beisert:2006ez,Kotikov:2007cy,Bajnok:2008qj,Beccaria:2009vt,Kniehl:2021ysp,Kniehl:2024tvd} and OPE coefficients~\cite{Eden:2012rr},
\begin{align}
\gamma_{2,\ell}(a)&=\sum_{i\ge1}a^i\gamma_i(\ell), \hspace{2em}
a_{2,\ell}(a) =\sum_{i\ge0}a^i a^{(i)}_{2,\ell}
=a_0(\ell)\left[1+\sum_{i\ge1}a^ir_i(\ell)\right],
\label{eq:twist-two-data}
\end{align}
where $a_0=a^{(0)}_{2,\ell}$ and $r_i=a^{(i)}_{2,\ell}/a_0$.  The four-loop resummation needs $\gamma_1,\ldots,\gamma_4$ and $r_1,r_2,r_3$.  Expanding the twist-two block through
$L_z^n,L_z^{n-1},L_z^{n-2},L_z^{n-3}$ gives
\begin{align}
\mathcal F^{(n)}(z,\bar z)
=z^3\sum_{\ell\ {\rm even}}\sum_{q=0}^{3}L_z^{\,n-q} P_{q}(\ell,\bar z) +O(z^4),
\label{eq:twist-two-kernel-source}
\end{align}
for example,
\begin{equation}
    P_0(\ell,\zb) =\frac{a^{(0)}_{2,\ell}\gamma_1^n}{2^n n!} \,k_{6+2\ell}(\bar z), \quad  P_1(\ell,\zb) = 
\left(
\frac{a^{(1)}_{2,\ell}\gamma_1^{n-1}}{2^{n-1}(n-1)!}
+\frac{a^{(0)}_{2,\ell}\gamma_2\gamma_1^{n-2}}{2^{n-1}(n-2)!}
\right)k_{6+2\ell}(\bar z)
+\frac{a^{(0)}_{2,\ell}\gamma_1^n}{2^n (n-1)!}\,
\partial_\ell k_{6+2\ell}(\bar z),
\end{equation}
where $q=0,1,2,3$ labels LL through N$^3$LL contributions and $P_{q}(\ell,\bar z)$ can be derived from the perturbative expansion.

\medskip
\noindent\emph{Large spin analysis.}
The spin derivative is a bookkeeping device for the large-spin tail of the even-spin sum. We first use the IBP method to move the derivatives from $k_{6+2\ell}(\zb)$ onto the OPE data and anomalous dimensions, and then replace the resulting derivative basis by fixed even-spin finite differences because we will consider large spin expansion. Thus this step is a formal summation-by-parts operation followed by a discrete shift-normal representation. The reduction has the schematic form
\begin{align}
P_q(\ell,\zb) \quad \longrightarrow \quad k_{6+2\ell}(\bar z)\,a_0(\ell)\sum_{s=0}^q\Delta_sB_{q,s}(\ell),
\label{eq:kernel-to-shift-basis}
\end{align}
where $s$ is the finite-difference shift index and $B_{q,s}$ is the coefficient relying $\ell$. The shift $E_\delta$ acts on the whole displayed coefficient, $E_\delta f(\ell)=f(\ell+\delta)$, including the derivatives of $a_0(\ell)$. The finite-difference stencils used in the calculation are
\begin{align}
\Delta_0=E_0,\qquad
\Delta_1=\frac{E_{-2}-E_2}{4},\qquad
\Delta_2=\frac{E_4-2E_0+E_{-4}}{16},
\qquad
\Delta_3=\frac{E_{-4}-2E_{-2}+2E_2-E_4}{16}.
\label{eq:shift-basis-examples}
\end{align}
This is the step that keeps shifted products such as $\rho(\ell\pm2)$, with $\rho=\partial_\ell\log a_0$, tied to the whole coefficient rather than to isolated factors. We can define the following notation $D_q^{\rm shift}$ to denote the same source before that large-spin expansion, written directly in the finite-difference shift basis,
\begin{align}
D_q^{\rm shift}(\ell)&=a_0(\ell)\sum_{s=0}^{q}\Delta_s B_{q,s}(\ell),
\nonumber\\
B_{q,0}&=H_{q,0}-\rho H_{q,1}
        +(\rho^2-\rho')H_{q,2}
        +(-\rho^3+3\rho\rho'-\rho'')H_{q,3},
\label{eq:shift-normal-sketch}
\end{align}
with analogous expressions for $B_{q,1},B_{q,2},B_{q,3}$.  In this formula all factors are evaluated at the common shifted spin argument before $\Delta_s$ acts, $\rho=\partial_\ell\log a_0$, and primes denote $\ell$-derivatives.  For example, at $q=2$ the nonzero coefficients are
\begin{align}
H_{2,0}&=
\frac{r_2\gamma_1^{n-2}}{2^{n-2}(n-2)!}
+\frac{r_1 \gamma_2\gamma_1^{n-3}}{2^{n-2}(n-3)!}
+\frac{\gamma_3\gamma_1^{n-3}}{2^{n-2}(n-3)!}
+\frac{\gamma_2^2\gamma_1^{n-4}}{2^{n-1}(n-4)!},
\nonumber\\
H_{2,1}&=
\frac{r_1\gamma_1^{n-1}}{2^{n-1}(n-2)!}
+\frac{(n-1)\gamma_2\gamma_1^{n-2}}{2^{n-1}(n-2)!},
\qquad
H_{2,2}=\frac{\gamma_1^n}{2^{n+1}(n-2)!}.
\label{eq:nnll-h-source}
\end{align}
All quantities in \Cref{eq:nnll-h-source} can be derived before the shift operators in \Cref{eq:shift-basis-examples} act.

After this basis reduction, we can impose the large-spin expansion on the OPE data and anomalous dimensions. We denote by $D_q^{\rm LS}(\ell)$ the large-spin expansion of $D_q^{\rm shift}(\ell)$ through order $J_2^{-1}$, with the logarithms required at N$^3$LL retained. A useful representative is
\begin{align}
\mathcal F^{(n)}(z,\bar z)
&=z^3\sum_{\ell\ {\rm even}} k_{6+2\ell}(\bar z)
\sum_{q=0}^{3}L_z^{\,n-q}D_q^{\rm LS}(\ell) +O(z^4),
\label{eq:direct-large-spin-sketch}
\end{align}
where $q=0,1,2,3$ labels LL through N$^3$LL. In the large spin expansion, we can use the twist-two conformal spin
\begin{align}
J_2=(\ell+2)(\ell+3),\qquad L_J=\log J_2+2\gamma_E .
\end{align}
The LL contribution in $z$ ($q=0$ case) already shows how LP and NLP information coexist in the same large-spin,
\begin{align}
D_0^{\rm LS}(\ell)=
\frac{a_0(\ell)}{2^n n!}
\left[L_J^n+\frac{n}{3J_2}L_J^{n-1}\right].
\label{eq:large-spin-example}
\end{align}
The $J_2^0$ term feeds LP, while the $J_2^{-1}$ term feeds NLP. The NLL contribution adds the expected $\zeta_2$ and $\zeta_3$ structures, for example
\begin{align*}
D_1^{\rm LS}(\ell)
&=\frac{a_0(\ell)}{2^n\Gamma(n)}
\left[-(n+1)\zeta_2L_J^{n-1}-3(n-1)\zeta_3L_J^{n-2}
+\frac{1}{J_2}
\left(
(n-1)\left(1-\frac{(n+1)\zeta_2}{3}\right)L_J^{n-2}
-(n-1)(n-2)\zeta_3L_J^{n-3}
\right) \right],
\label{eq:large-spin-nll-example}
\end{align*}
and the NNLL and N$^3$LL terms are obtained from the same construction. The notation $D_q^{\rm LS}$ refers to this large-spin expansion $D_q^{\rm LS}(\ell)=\lim_{\ell \to \infty}D_q^{\rm shift}(\ell)$ up to the order $J_2^{-1}$. 

\medskip
\noindent\emph{Twist conformal block basis.}
The even large-spin monomials are then lifted to the shifted TCB basis~\cite{Alday:2016njk}. For twist two contributions we have
\begin{equation}\label{eq:tcb-definition}
H_2^{(m,i)}(z,\bar z)
=\sum_{\ell\ {\rm even}}a_0(\ell)
\frac{\log^i J_2}{J_2^m}\, G_{6,\ell}(z,\bar z),
\end{equation}
where $m=0$ is the LP TCB sector and $m=1$ is the NLP TCB sector retained here; $i$ is a non-negative logarithm index.  Since the large-spin source uses $L_J=\log J_2+2\gamma_E$, the wrapped basis absorbs the Euler-constant shift,
\begin{align}
\mathcal L\!\left[c\,J_2^{-m}L_J^p\right]
=c\,\mathbb H_p^{(m)}, \hspace{3em}
\mathbb H_p^{(m)}
=\sum_{i=0}^{p}\binom{p}{i}(2\gamma_E)^i H_2^{(m,p-i)} .
\label{eq:wrapped-tcb-map}
\end{align}
In the small-$\epsilon$ region the singular part factorizes as
\begin{align}
H_2^{(m,i)}(z,\bar z)
=\frac{z\,k_4(z)}{\bar z-z}\bar H_2^{(m,i)}(\bar z)
+H_{2,{\rm reg}}^{(m,i)}(z,\bar z).
\end{align}
The bar denotes the essential $\bar{z}$ dependence of the twist conformal block. For the purpose of this work, we only keep the $\epsilon^{-1}$ and $\epsilon^0$ terms in $\bar H_2^{(m,i)}$, which can feed the final resummation of EEC at LP and NLP; terms contribute to $O(y)$ are discarded. With this rule, we derive the explicit expansion as 
\begin{align}\label{eq:H0i}
\bar H_2^{(0,i)}
={}&(-1)^i\Bigg\{
\frac{1}{\epsilon}\Bigg[
\frac{L_\epsilon^i}{2}
+i\gamma_E L_\epsilon^{i-1}
+(i-1)_2\left(\gamma_E^2+\frac{\pi^2}{12}\right)L_\epsilon^{i-2} + (i-2)_3\left(
\frac{2\gamma_E^3}{3}+\frac{\gamma_E\pi^2}{6}+\frac{\zeta_3}{3}
\right)L_\epsilon^{i-3}
\nonumber\\[-2pt]
&\hspace{5.2em}
+(i-3)_4\left(
\frac{\gamma_E^4}{3}+\frac{\gamma_E^2\pi^2}{6}
+\frac{7\pi^4}{720}+\frac{2\gamma_E\zeta_3}{3}
\right)L_\epsilon^{i-4}
\Bigg]
+\frac{L_\epsilon^{i+1}}{6(i+1)}
+\left(\frac{\gamma_E}{3}-1\right)L_\epsilon^i
\nonumber\\[-2pt]
&\hspace{2.6em}
+i\left(
\frac13-2\gamma_E+\frac{\gamma_E^2}{3}+\frac{\pi^2}{36}
\right)L_\epsilon^{i-1}
+(i-1)_2 \left(
\frac{2\gamma_E}{3}-2\gamma_E^2+\frac{2\gamma_E^3}{9}
-\frac{\pi^2}{6}+\frac{\gamma_E\pi^2}{18}+\frac{\zeta_3}{9}
\right)L_\epsilon^{i-2}
\nonumber\\[-2pt]
&\hspace{2.6em}
+(i-2)_3\left(
\frac{2\gamma_E^2}{3}-\frac{4\gamma_E^3}{3}+\frac{\gamma_E^4}{9}
+\frac{\pi^2}{18}-\frac{\gamma_E\pi^2}{3}
+\frac{\gamma_E^2\pi^2}{18}+\frac{7\pi^4}{2160}
-\frac{2\zeta_3}{3}+\frac{2\gamma_E\zeta_3}{9}
\right)L_\epsilon^{i-3} 
\Bigg\}+\cdots,
\end{align}
and 
\begin{align}
\bar H_2^{(1,i)}
={}&(-1)^i\Bigg[
\frac{L_\epsilon^{i+2}}{2(i+1)_2}
+\frac{\gamma_E L_\epsilon^{i+1}}{i+1}
+\left(\gamma_E^2+\frac{\pi^2}{12}\right)L_\epsilon^i
+i\left(
\frac{2\gamma_E^3}{3}+\frac{\gamma_E\pi^2}{6}+\frac{\zeta_3}{3}
\right)L_\epsilon^{i-1}
\nonumber\\[-2pt]
&\hspace{2.6em}
+(i-1)_2\left(
\frac{\gamma_E^4}{3}+\frac{\gamma_E^2\pi^2}{6}
+\frac{7\pi^4}{720}+\frac{2\gamma_E\zeta_3}{3}
\right)L_\epsilon^{i-2}
\nonumber\\[-2pt]
&\hspace{2.6em}
+(i-2)_3\left(
\frac{2\gamma_E^5}{15}+\frac{\gamma_E^3\pi^2}{9}
+\frac{7\gamma_E\pi^4}{360}+\frac{2\gamma_E^2\zeta_3}{3}
+\frac{\pi^2\zeta_3}{18}+\frac{\zeta_5}{5}
\right)L_\epsilon^{i-3}
\Bigg]+\cdots .
\label{eq:H1i}
\end{align}
Here $\epsilon=1-\bar z$, $(a)_b=\Gamma(a+b)/\Gamma(a)$ represents the Pochhammer symbol, and the ellipses stand for terms either less singular in $\epsilon$ than the LP and NLP tower or lower in the logarithmic depth so that do not contribute to N$^3$LL resummation. The method for deriving these results can be found in~\cite{Henriksson:2017eej}. 

\medskip
\noindent\emph{Crossing symmetry of the local correlator.}
To avoid the complication of twist-4 CFT data, we make full use of the crossing symmetry of the local correlator: $\Phi(u,v)=\Phi(v,u)$.
Let us define
\begin{align}
\frac{\mathcal F^{(n)}(z,\bar z)}{z^3}
=\sum_{q=0}^{3}L_z^{\,n-q}\mathfrak F_{q,n}(\epsilon,L_\epsilon),
\label{eq:f-level-definition}
\end{align}
where $\mathfrak F_{q,n}$ is sourced from the wrapped TCB replacement
$\mathbb H_p^{(m)}\mapsto\mathcal R_p^{(m)}$.  For example, the NNLL in $z$ term is
\begin{align}
\mathfrak F_{2,n}
=\frac{1}{2^{n+1}\Gamma(n-1)}
\left[
\sum_{r'=0}^{4}S_{2,r'}^{\rm LP}(n)\mathcal R_{n-r'}^{(0)}
+\sum_{r'=1}^{5}S_{2,r'}^{\rm NLP}(n)\mathcal R_{n-r'}^{(1)}
\right],
\label{eq:nnll-f-layer}
\end{align}
where the $S_{2,r'}$ are the shifted-TCB coefficients.  The retained
wrapped block is
\begin{align}
\mathcal R_p^{(m)}
=\left[
\sum_{i=0}^{p}\binom{p}{i}(2\gamma_E)^i
\bar H_2^{(m,p-i)}(\bar z)
\right]_{\text{keep } \mathcal{O}(\epsilon^{-1}) \text{ and } \mathcal{O}(\epsilon^{0})}.
\label{eq:Rpm-definition}
\end{align}
For $m=0$, we keep the LP and NLP terms in $\epsilon$ ($\mathcal{O}(\epsilon^{-1})$ and $\mathcal{O}(\epsilon^{0})$) ; for $m=1$, we keep the $\mathcal{O}(\epsilon^{0})$ contribution. For example, $\mathcal R_0^{(0)}=1/(2\epsilon)+L_\epsilon/6+(\gamma_E-3)/3+\cdots$ and $\mathcal R_0^{(1)}=L_\epsilon^2/4+\gamma_E L_\epsilon+\cdots$, where the ellipses are the power suppressed terms. 

We can write the N$^q$LL term as
\begin{align}
\mathcal K_q(L_z,L_\epsilon,\epsilon)
&=L_z^{\,n-q}\mathfrak F_{q,n}
=\frac{A_q(L_z,L_\epsilon)}{\epsilon}
+B_q(L_z,L_\epsilon)+O(\epsilon).
\label{eq:nnll-k-layer}
\end{align}
Then we convert to crossing symmetric function $\Phi$ using $\mathcal F^{(n)}=(u^3/v)\Phi^{(n)}$, $v=\epsilon$, and $L_z=L_u+\epsilon$. The finite NLP correction induced by this rewrite is
\begin{align}
p_q(L_u,L_v)
&=\left[\partial_{L_z}A_q+B_q+3A_q
\right]_{L_z=L_u,\ L_\epsilon=L_v},
\nonumber\\
\Phi_q^{\rm full}
&=\frac12\left(A_q+A_q^\times\right)
+v\,p_q+u\,p_q^\times,
\qquad X^\times(L_u,L_v)=X(L_v,L_u).
\label{eq:nnll-phi-layer}
\end{align}
The derivative term comes from $L_z=L_u+\epsilon$, while the $3A_q$ term comes from the prefactor $(1-\epsilon)^{-3}$. The three sector labels $(j_u,j_v)=(0,0),(1,0),(0,1)$ denote the LP sector ($u^0 v^0$) and the two crossed NLP sectors ($u^1 v^0$ and ($u^0 v^1$)). Thus $\Phi_q^{\rm full}$ is the full crossed support before Mellin
projection.

\medskip
\noindent\emph{EEC from Mellin projection.} Starting from the Mellin representation of EEC and its kernel we have
$$
{\text {EEC}}(y)=\int \frac{dj_1dj_2}{(2\pi i)^2}\,M(j_1,j_2)
 K(j_1,j_2;y) \, , \hspace{0.5em} K(j_1,j_2;y)= \frac{4\Gamma(1-j_1-j_2)}
{\Gamma(j_1+j_2)\left[\Gamma(1-j_1)\Gamma(1-j_2)\right]^2}
\left(\frac{y}{1-y}\right)^{j_1+j_2}.
$$
Thus a monomial $u^{j_u}v^{j_v}L_u^\alpha L_v^\beta$ is mapped to
\begin{align}
\mathcal P_{j_u,j_v}^{\alpha,\beta}(y)
&=\alpha!\,\beta!\,
\operatorname*{Res}_{j_1=j_u}
\operatorname*{Res}_{j_2=j_v}
\frac{K(j_1,j_2;y)}
{4y(1-y)^2(j_1-j_u)^{\alpha+1}(j_2-j_v)^{\beta+1}},
\label{eq:projection-residue-sketch}
\end{align}
with pole locations $(j_u,j_v)=(0,0),(1,0),(0,1)$.  If the N$^q$LL contribution of the correlator is
$\Phi_q^{(n)}=\sum a^{(q)}_{j_u,j_v,\alpha,\beta}
u^{j_u}v^{j_v}L_u^\alpha L_v^\beta$ and set $q=k-1$, the compact extraction formula is
\begin{align}
C^{\rm LP}_{k}(n)
&=
\left[\sum_{\alpha,\beta}a^{(k-1)}_{0,0,\alpha,\beta}(n)
\mathcal P_{0,0}^{\alpha,\beta}(y)\right]_{y^{-1} L_y^{2n-k}}, \\
C^{\rm NLP}_{k}(n)
&=\left[
\sum_{\alpha,\beta}a^{(k-1)}_{0,0,\alpha,\beta}(n)
\mathcal P_{0,0}^{\alpha,\beta}(y)
+\sum_{\alpha,\beta}a^{(k-1)}_{1,0,\alpha,\beta}(n)
\mathcal P_{1,0}^{\alpha,\beta}(y)
+\sum_{\alpha,\beta}a^{(k-1)}_{0,1,\alpha,\beta}(n)
\mathcal P_{0,1}^{\alpha,\beta}(y)
\right]_{y^0L_y^{2n-k}}.
\label{eq:coeff-extraction-general}
\end{align}

\medskip
\noindent\emph{Results.} The formulas below are the generic large-spin tail coefficients. With the normalization of \Cref{eq:b2b-coeff-map}, the LP towers are
\begin{align}\label{eq:lp-tower-coefficients}
C^{\rm LP}_1(n)&=\frac{(-1)^n}{2^{n}\Gamma(n)}, \quad C^{\rm LP}_3(n)=\frac{(-1)^n(n+1)\pi^2}
{3\,2^{n+1}\Gamma(n-1)},  \quad  C^{\rm LP}_4(n)=\frac{(-1)^n(16n^2-50n+39)\zeta_3}
{3\,2^{n}\Gamma(n-1)} .
\end{align}
Here $C^{\rm LP}_2(n)=0$.  The generic coefficients at NLP towers are
\begin{align}\label{eq:nlp-tower-coefficients}
C^{\rm NLP}_1(n)&=\frac{(-1)^n n}{2^{n-1}(2n-1)\Gamma(n)}, \hspace{9em} C^{\rm NLP}_2(n)=\frac{(-1)^n(7n-5)}
{3\,2^{n+1}\Gamma(n)}, \nonumber\\
C^{\rm NLP}_3(n)&=
\frac{(-1)^n\left(\pi^2(2n^2+n-2)+12\right)}
{3\,2^{n+1}(2n-3)\Gamma(n-1)}, \hspace{3em}
C^{\rm NLP}_4(n) = \frac{(-1)^n}{2^{n-3}\Gamma(n-1)}
\left(\zeta_3 Q_1(n)+\pi^2 Q_2(n)-\frac18\right) . 
\end{align}
Here $N_c$ is the rank of the gauge group. The two polynomials appearing in the N$^3$LL NLP coefficient are
\begin{align*}
Q_1(n)=\frac{16n^2-39n+15}{24}
+\frac{3n-9}{2N_c^2},\quad Q_2(n)=\frac{7n^2-2n-3}{288}
+\frac{n-3}{8N_c^2}.
\end{align*}
\begin{figure}[t!]
    \centering
    \includegraphics[width=0.8\linewidth]{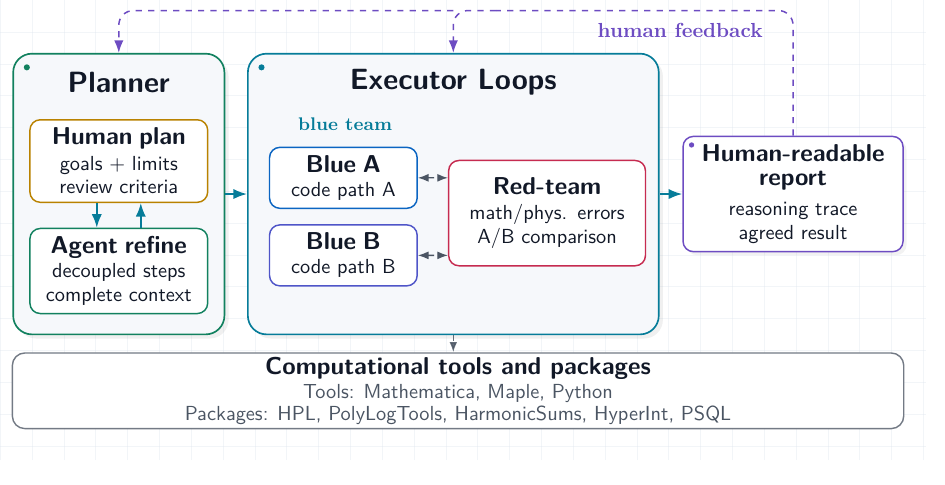}
    \caption{\justifying 
    Human-in-the-Loop agentic workflow. The computation is decomposed by a planner, executed through redundant Blue team paths, reviewed by a Red-team validator, and returned to the human researchers as a compact report for iterative feedback.}
    \label{fig:workflow}
\end{figure}
The fixed-order endpoint expansion provides boundary coefficients that are not determined by large-spin data. One needs to expand the fixed-order EEC$^{(n)}$ to modify the boundary condition. We collect the expansion of fixed-order EEC$^{(n)}$ \cite{Belitsky:2013ofa,Henn:2019gkr} around $y\to 0$ up to three-loop ($n=3$) here
\begin{align*}
\mathrm{EEC}^{(1)}
& =\frac{1}{y}\left( -\frac{L_y}{2} \right)  -{L_y}  +\mathcal O(y),
\\
\mathrm{EEC}^{(2)}
& =\frac{1}{y}\left(
\frac{L_y^3}{4}
+\frac{\pi^2}{8}L_y
+\frac{\zeta_3}{4}
+\cdots
\right) +\left(
\frac{L_y^3}{3}
+\frac{3}{8}L_y^2
+\frac{3+2\pi^2}{6} L_y
+\frac{-36+\pi^2(-1+18\log 2)+15\zeta_3}{24} 
+\cdots
\right)
+\mathcal O(y), \\
\mathrm{EEC}^{(3)}
&=\frac{1}{y}\left(
-\frac{L_y^5}{16}
-\frac{\pi^2}{12}L_y^3
-\frac{11\zeta_3}{8}L_y^2
+\cdots
\right)
+\left(
-\frac{3L_y^5}{40}
-\frac{L_y^4}{6}
-\frac{12+19\pi^2}{144}L_y^3
+\frac{2-3\pi^2-28\zeta_3}{16}L_y^2
+\cdots
\right)
+\mathcal O(y),
\end{align*}
substituting these conditions in the generic
N$^3$LL expressions, we fix
\begin{equation}
    C^{\rm NLP}_2(1)=0 , \hspace{3em}  C^{\rm NLP}_4(2)=  \frac{-36+\pi^2(-1+18\log 2)+15\zeta_3}{24} .
\end{equation}
The remaining terms are in exact agreement with our prediction of generic LP and NLP coefficients, which independently checks the correctness of the formulas derived from the twist conformal block resummation. Substituting the generic and boundary coefficients into the logarithmic towers and summing over $n$ gives
\begin{align}
    {\rm EEC} (y) =  \frac{1}{y} \mathcal G_{y} F(L_y,a) +  G(L_y,a) + \mathcal{O}(y).
\end{align}
where $\mathcal G_y=e^{-aL^2_y/2}$. The first term collects the leading-power Sudakov tower, while $G(L_y,a)$ involves the next-to-leading power contribution. For the leading-power part we find
\begin{align*}
F(L_y,a)=
&-\frac{1}{2} a L_y+\frac{1}{8}\pi^2 a^2 L_y+\frac{1}{4} a^2\zeta_3 -\frac{1}{48}\pi^2 a^3 L_y^3-\frac{5}{4} a^3\zeta_3 L_y^2+\frac{1}{3} a^4\zeta_3 L_y^4+\cdots .
\end{align*}
The NLP function $G(L_y,a)$ is given in \Cref{eq:NLPfunG} of the main text. The whole N$^3$LL EEC calculation was implemented with GPT-5.5 in the human-in-the-loop (HITL) workflow described by~\Cref{fig:workflow} .

\end{document}